%% file: paper.tex
\pdfoutput=1

\documentclass[12pt,a4paper]{article}

\usepackage{ifthen} 
\newboolean{pdflatex}
\setboolean{pdflatex}{true} 

\newboolean{articletitles}
\setboolean{articletitles}{true} 

\newboolean{uprightparticles}
\setboolean{uprightparticles}{false} 

\newboolean{inbibliography}
\setboolean{inbibliography}{false} 

\usepackage{multirow}
\usepackage{booktabs}
\def\paperauthors{LHCb collaboration} 
\def\paperasciititle{Measurement of the B^{\pm} production cross-section in pp collisions at sqrt(s) = 7 and 13 TeV} 
\def\papertitle{ Measurement of the \Bpm production cross-section in $pp$ collisions at $\sqrt{s}=$~7 and 13\tev} 
\def\paperkeywords{{High Energy Physics}, {LHCb}} 
\def\papercopyright{CERN on behalf of the LHCb collaboration}
\def\paperlicence{CC-BY-4.0}
\def\paperlicenceurl{https://creativecommons.org/licenses/by/4.0/}

\input{preamble}
\usepackage{longtable} 

\newcommand{\jpsimumu}{\nobreak{\ensuremath{ \jpsi \rightarrow \mup \mun }}}
\newcommand{\bujpsiks}{\nobreak{\ensuremath{\Bpm \rightarrow \jpsi \Kpm }}}
\newcommand{\ratiores}{\nobreak{\ensuremath{ 2.02\pm0.02\stat\pm0.12\syst }}}
\def\bplusplotwidth{0.48\textwidth}
\def\bplusoneplotwidth{0.8\textwidth}

\begin{document}

\renewcommand{\thefootnote}{\fnsymbol{footnote}}
\setcounter{footnote}{1}

\input{title-LHCb-PAPER}


\renewcommand{\thefootnote}{\arabic{footnote}}
\setcounter{footnote}{0}

\cleardoublepage


\pagestyle{plain} 
\setcounter{page}{1}
\pagenumbering{arabic}


\section{Introduction}
Precise measurements of the production cross-section of \Bpm~mesons
in $pp$ collisions provide important
tests of perturbative quantum chromodynamics (QCD) calculations,
particularly of the state-of-the-art calculations based on the fixed next-to-leading order (NLO) QCD
with next-to-leading logarithm (NLL) large transverse momentum resummation (FONLL) 
approach{~\cite{Cacciari:1998it,Cacciari:2001td}}.
The FONLL calculations are accurate to the full NLO level at moderate \pt values,
and to the NLL level at high \pt. The FONLL predictions are then achieved by properly merging a 
`massless' resummed approach, valid in the high \pt region, with a full
massive fixed-order calculation, reliable in the small \pt region.
The ratio of cross-sections between different centre-of-mass energies
is of particular interest due to cancellations that occur in the theoretical and experimental uncertainties. 
Uncertainties arising from assumptions about the values of the FONLL
parameters largely cancel in the ratio. 
Experimentally, uncertainties due to factors such as the branching fractions of decays
and the $b$-quark fragmentation fractions~\cite{LHCb-PAPER-2011-018} to specific hadrons are highly
correlated at different beam energies and their effect is much reduced
in the ratio.

Previous measurements of \Bpm production have been performed in different kinematic regions
at the centre-of-mass energy $\sqrt{s}=$~7\tev by several experiments at the Large Hadron Collider.
The \cms collaboration reported the integrated and differential \Bpm
production cross-sections in the range ${\pt > 5\gevc}$ and ${|y|<2.4}${~\cite{Khachatryan:2011mk,Khachatryan:2016csy}},
where \pt and $y$ are the component of the momentum transverse to the beam line and the rapidity of the \Bpm~mesons, respectively.
The \atlas collaboration measured the production cross-sections
in the range ${9 <\pt <120\gevc}$ and ${|y|<2.25}$~\cite{ATLAS:2013cia}.
The \lhcb collaboration measured the integrated and
differential cross-sections for \Bpm with ${0<\pt<40\gevc}$ and
${2.0<y<4.5}$ using a data sample collected in 2010
that corresponds to an integrated luminosity of $35\invpb$~\cite{LHCb-PAPER-2011-043}.
This result was later updated using a data sample collected in early 2011,
corresponding to an integrated luminosity of ${362\invpb}$~\cite{LHCb-PAPER-2013-004}.

This article updates the previous LHCb results using a larger data sample collected in 2011
with the \lhcb experiment at $\sqrt{s}=$~7\tev and 
corresponding to an integrated luminosity of ${1.0\invfb}$. 
The first measurements of the integrated and differential cross-sections
of \Bpm~mesons at $\sqrt{s}=$~13\tev are also presented, using a data sample
collected in 2015 and corresponding to an integrated luminosity of ${0.3\invfb}$.
Following the previous \lhcb measurements{~\cite{LHCb-PAPER-2011-043, LHCb-PAPER-2013-004}},
the \Bpm~mesons are reconstructed in the \bujpsiks~mode followed by \jpsimumu~and 
the production cross-sections are measured in the range ${0<\pt<40\gevc}$ and ${2.0<y<4.5}$.
The ratio of the cross-sections in the $13\tev$ and $7\tev$ data is also measured 
as a function of $\pt$ and $y$.

\section{Event selection}
The \lhcb detector{~\cite{Alves:2008zz,LHCb-DP-2014-002}} is a single-arm forward
spectrometer covering the \mbox{pseudorapidity} range ${2<\eta <5}$,
designed for the study of particles containing \bquark or \cquark~quarks. 
The detector includes a high-precision tracking system
consisting of a silicon-strip vertex detector surrounding the $pp$
interaction region, a large-area silicon-strip detector located
upstream of a dipole magnet with a bending power of about
$4{\mathrm{\,Tm}}$, and three stations of silicon-strip detectors and straw
drift tubes placed downstream of the magnet.
The tracking system provides a measurement of momentum, \ptot, of charged particles with
a relative uncertainty that varies from 0.5\% at low momentum to 1.0\% at 200\gevc.
The minimum distance of a track to a primary vertex, the impact parameter,
is measured with a resolution of ${(15+29/\pt)\mum}$,
where \pt is in units of\,\gevc.
Different types of charged hadrons are distinguished using information
from two ring-imaging Cherenkov detectors.
Photons, electrons and hadrons are identified by
a calorimeter system consisting of scintillating-pad (SPD) and preshower detectors, 
an electromagnetic calorimeter and a hadronic calorimeter.
Muons are identified by a system composed of alternating layers of iron and multiwire
proportional chambers.

The online event selection is performed by a trigger,
which consists of a hardware stage (L0),
based on information from the calorimeter and muon systems,
followed by a two-stage software-based high-level trigger 
(HLT1, HLT2)~\cite{LHCb-DP-2012-004},
where the HLT1 stage uses partial event reconstruction to 
reduce the rate and the HLT2 stage applies a full event reconstruction. 
At the hardware stage, events are required to have a dimuon candidate with large \pt,
while at the software stage the dimuon invariant mass is required to be consistent with the known
\jpsi mass~\cite{PDG2016}.
Finally, a set of global event cuts (GEC) is applied in order to
prevent high-multiplicity events from dominating the processing time of the software trigger,
which includes the requirement that the number of hits in the SPD subdetector should be less than 900.

Simulated events are used to optimise the selection, determine some of the efficiencies
and estimate the background contamination.
The simulation is based on the {\sc Pythia8}
generator~\cite{Sjostrand:2007gs}
with a specific \lhcb configuration~\cite{LHCb-PROC-2010-056}.
Decays of hadrons are described by \evtgen~\cite{Lange:2001uf},
in which final-state radiation is generated using \photos~\cite{Golonka:2005pn}.
The interaction of the generated particles with the detector, and its response,
are implemented using the \geant toolkit{~\cite{Allison:2006ve, *Agostinelli:2002hh}},
as described in Ref.~\cite{LHCb-PROC-2011-006}.

The \Bpm candidates are made by combining \jpsi and \Kpm candidates.
The offline event selection forms \jpsi candidates using pairs of
 muons with opposite charge.
The muon candidates must have ${\pt>700\mevc}$, and satisfy track-reconstruction quality 
and particle identification (PID) requirements.
The muon pair is required to be consistent with
originating from a common vertex~\cite{Hulsbergen:2005pu}
and have an invariant mass, $M(\mu\mu)$, within the range ${3.04<M(\mu\mu)<3.14\gevcc}$.
The \Kpm candidates are required to have $\pt$
greater than 500\mevc and satisfy track-reconstruction quality requirements.
No PID requirement is applied to select the kaon, as the only topologically similar decay,
$\Bpm\to\jpsi\pipm$, is Cabibbo suppressed.

The three tracks in the final state of the \Bpm decay are required to form a common vertex
with a good vertex-fit quality.
In order to suppress background due to the random 
combination of particles produced in the $pp$ interactions,
the \Bpm candidates are required to have a decay time larger than 0.3\ps.
Finally, \Bpm candidates with ${0<\pt<40\gevc}$ and
${2.0<y<4.5}$ are selected for subsequent analysis.

\section{Cross-section determination}
The differential production cross-section of \Bpm~mesons is measured as a function
of \pt and $y$ using
\begin{eqnarray}
\label{eq:xsec}
 \frac{\deriv^{2}\sigma}{\deriv y\deriv\pt} = \frac{N_{\Bpm}}{{\lum}\times \eps_{\mathrm{tot}}\times {\BR} \left( \Bpm\to\jpsi\Kpm\right) \times {\BR} \left(\jpsi\to\mumu\right) \times \Delta y \times \Delta\pt },
\end{eqnarray}
where {$N_{\Bpm}$ is the number of reconstructed \bujpsiks~candidates in a given
$\text{(\pt, $y$)}$ bin after background subtraction, \lum is the integrated luminosity,
$\eps_{\mathrm{tot}}$ is the bin-dependent total efficiency,
${\BR} \left( \Bpm\rightarrow \jpsi \Kpm\right)$
is the branching fraction of $\Bpm$ decays to $\jpsi\Kpm${~\cite{Abe:2002rc,Aubert:2004rz}},
${\BR} \left( \jpsi \rightarrow \mumu\right)$ is the branching fraction of $\jpsi$
decays to $\mumu$}~\cite{PDG2016}, and $\Delta y$ and $\Delta \pt$ are the bin widths in $y$ and
$\pt$. 
The value of ${\BR} \left( \Bpm\rightarrow \jpsi \Kpm\right)$ is calculated to be ${(1.044 \pm 0.040)\times10^{-3}}$ by combining the 
two exclusive measurements from the \belle~\cite{Abe:2002rc} and \babar~\cite{Aubert:2004rz} collaborations, 
under the assumption that only the uncertainty for $\jpsi\to\mumu$ branching fraction is correlated. 

\begin{figure}
\centering
 \includegraphics[width=\bplusplotwidth]{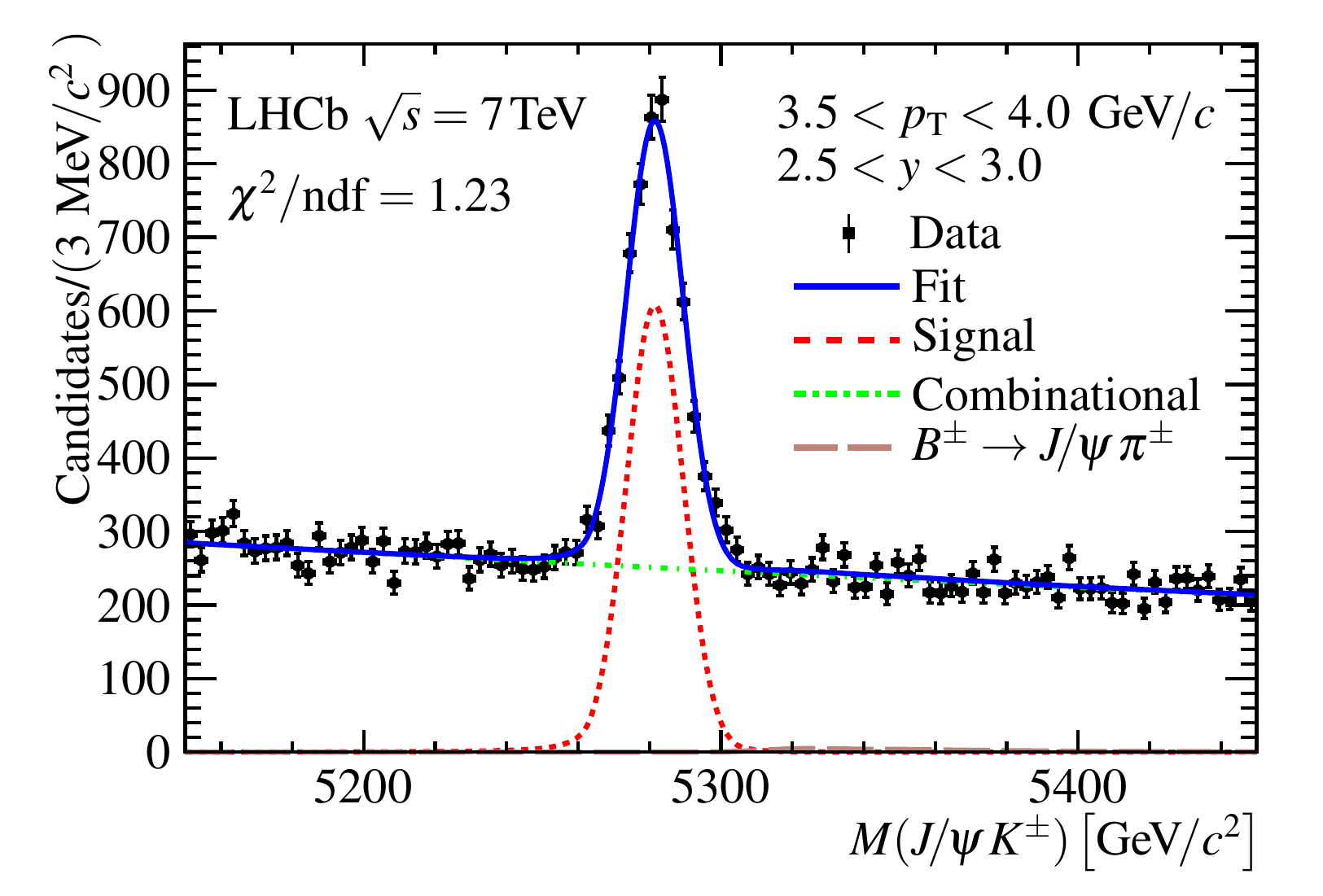}
 \includegraphics[width=\bplusplotwidth]{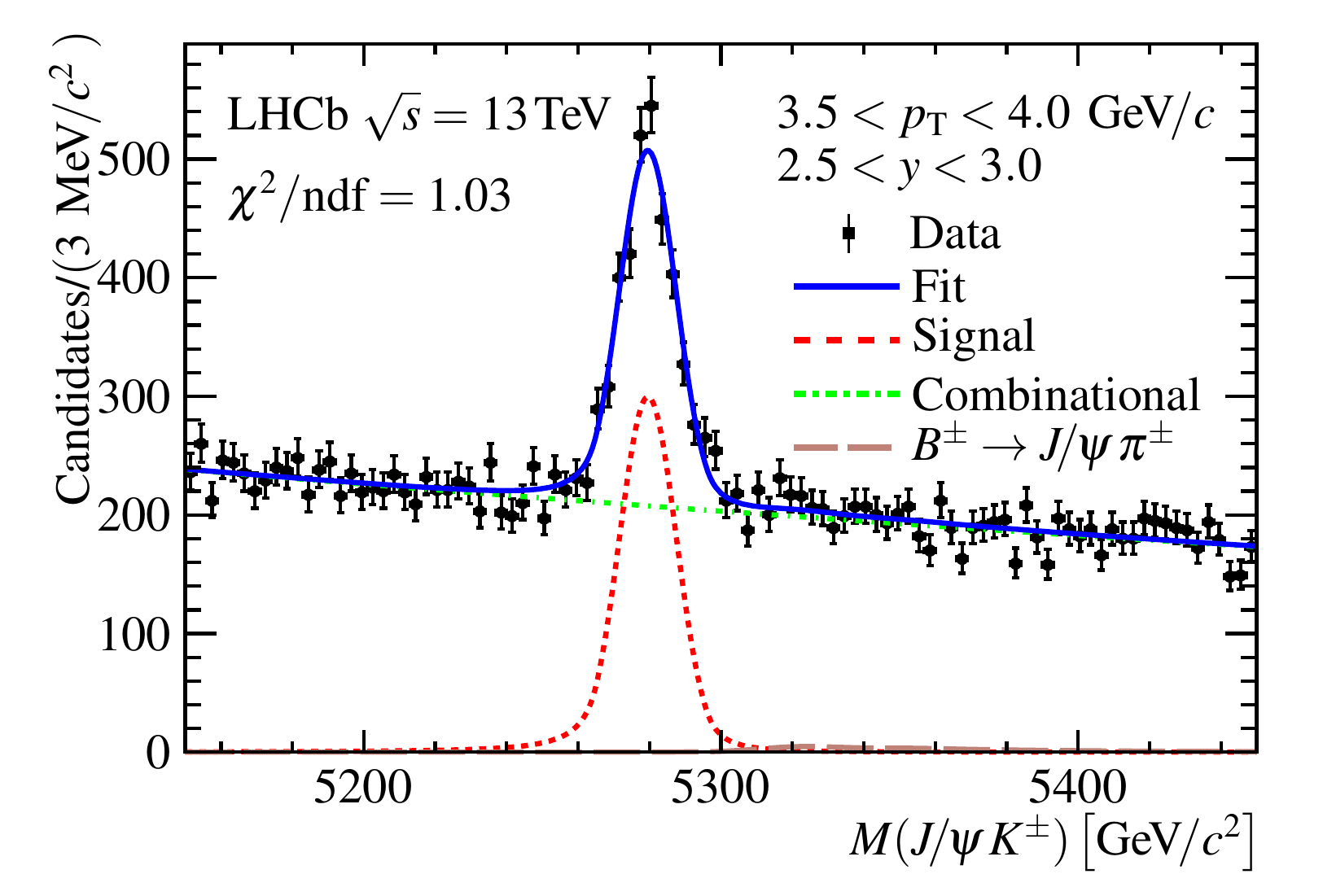}
\caption{Invariant mass distributions of \Bpm candidates in the range
${3.5<\pt<4.0\gevc}$ and ${2.5<y<3.0}$ using (left) $7\protect\tev$ and (right) $13\protect\tev$ data.
The black points are the number of selected candidates in each bin,
the blue curve represents the fit result, the red-dotted line
represents the $\Bpm\rightarrow \jpsi \Kpm$ signal, and the green- and brown-dashed lines
are contributions from the combinatorial and the Cabibbo-suppressed backgrounds.
The Cabibbo-suppressed background contribution $\Bpm\rightarrow \jpsi \pipm$ is only just visible at masses above the signal peak.}
\label{fig:mass_fit}
\end{figure}

The yield of \bujpsiks~decays in each (\pt, $y$) bin is obtained independently
by fitting the invariant mass distribution of the \Bpm candidates, $M(\jpsi\Kpm)$,
in the interval ${5150<M(\jpsi\Kpm)<5450\mevcc}$,
using an extended unbinned maximum likelihood fit.
The $M(\jpsi \Kpm)$ distribution is described by a probability density function (PDF)
consisting of the following three components: a modified Crystal Ball (CB) function~\cite{Skwarnicki:1986xj} 
to model the signal,
an exponential function to model the combinatorial background,
and a double CB function to model the contamination from
the Cabibbo suppressed decay $B^{\pm}\rightarrow \jpsi \pipm$,
where the charged pion is misidentified as a kaon.
The modified CB function used for the signal component has tails on both the low- and
the high-mass side of the peak, which are described by separate parameters.
In each bin, the tail parameters are determined from simulation, leaving the mean and width 
as free parameters.
The shape of the misidentification background is obtained using
simulated $\Bpm\rightarrow \jpsi \pipm$ decays that satisfy
the selection criteria for the decay $\Bpm\rightarrow \jpsi \Kpm$,
and the yield of the misidentification background is determined according to
the branching fraction ratio ${{\cal{B}}(\Bpm\to\jpsi\pi^{\pm})/ {\cal{B}}(\Bpm\to\jpsi K^{\pm})}$ from Ref.~\cite{PDG2016}.
Figure~\ref{fig:mass_fit} shows, as an example, the invariant mass distribution
 of the \Bpm candidates in the range ${3.5<\pt<4.0\gevc}$ and ${2.5<y<3.0}$.

The total efficiency, $\eps_{\text{tot}}$, is the product of several efficiencies and 
can be written as
\begin{eqnarray}
\label{eq:eff}
\varepsilon_{\text{tot}} = \varepsilon_{\text{acc}}\times\varepsilon_{\text{reco\&sel}}\times\varepsilon_{\text{PID}}\times\varepsilon_{\text{track}}
\times\varepsilon_{\text{trigger}}\times\varepsilon_{\text{GEC}}\,.
\end{eqnarray}
The acceptance factor, $\varepsilon_{\text{acc}}$, is the fraction of 
signal with all final-state particles within the fiducial region of the detector
acceptance, and is calculated from simulation, as shown in Appendix~\ref{app:acc}.
The efficiency of the particle reconstruction and event selection,
$\varepsilon_{\text{reco\&sel}}$, is also determined from simulation.
The efficiency of identifying the two muons in the final state, $\varepsilon_{\text{PID}}$,
and the track finding efficiency, $\varepsilon_{\text{track}}$,
are measured using a tag-and-probe method{~\cite{Lupton:2134057,LHCb-DP-2013-002}}
on a control data sample of $\jpsi\rightarrow \mumu$ decays.
The trigger efficiency, $\varepsilon_{\text{trigger}}$, is estimated in two parts, 
which are $\varepsilon_{\text{L0\&HLT1}}$ and $\varepsilon_{\text{HLT2}}$. 
The $\varepsilon_{\text{L0\&HLT1}}$ efficiency is evaluated by
estimating the fraction of events in a trigger-unbiased data sample
that satisfy the trigger requirements, and
the $\varepsilon_{\text{HLT2}}$ efficiency is evaluated using simulated signal events, as 
the effects of HLT2 trigger are well modelled. 
The GEC efficiency, $\varepsilon_{\text{GEC}}$, is 
measured to be ${(99.2\pm 0.1)}$\% for the 7\tev and ${(99.3\pm 0.1)}$\% for the 13\tev data sample,
and is independent of \pt and $y$.
It is extracted by fitting the SPD multiplicity distribution and extrapolating
the function to determine the fraction of events that are accepted.
The measured $\varepsilon_{\text{tot}}$ are tabulated in Appendix~\ref{app:eff}.
\section{Systematic uncertainties}
Sources of systematic uncertainty associated with
the determination of the
luminosity, branching fractions, signal yields, efficiencies,
along with their effects on the integrated cross-section measurements,
are summarised in Table~\ref{tab:uncer_tot}. 
The total systematic uncertainty is obtained from the sum in 
quadrature of all contributions.
Several uncertainties have been reduced in the {$\sqrt{s}=$ 13\tev} measurement, 
due to a larger simulation sample and a better understanding of the efficiency.

\begin{table}[t]
\begin{center}
\caption{\label{tab:uncer_tot} Summary of relative systematic uncertainties 
on the integrated production cross-sections at {$\sqrt{s}=$ 7} and $13\protect\tev$, and
 the ratio of the cross-sections $R(13\protect\tev/7\protect\tev)$.}
\begin{tabular}{lccc}
\hline
\hline
\multirow{2}{*}{Sources} & \multicolumn{3}{c}{Uncertainty (\%)} \\ 
 & 7\tev & 13\tev & $R(13\tev/7\tev)$\\ \hline
Luminosity          & 1.7  & 3.9   & 3.4\\
Branching fractions & 3.9  & 3.9   & 0.0\\
Binning             & 2.6  & 2.7   & 0.0\\
Mass fits           & 2.7  & 1.3   & 1.5\\
Acceptance          & 0.2  & 0.1   & 0.2\\
Reconstruction      & 0.1  & 0.1   & 0.2\\
Track               & 1.6  & 2.6   & 1.0\\
PID                 & 0.4  & 0.1   & 0.4\\
Trigger             & 3.5  & 2.6   & 4.4\\
GEC                 & 0.7  & 0.7   & 1.0\\
Selection           & 1.0  & 1.1  & 0.1\\
Weighting           & 0.2  & 0.2   & 0.3 \\ \hline
Total               & 7.0  & 7.4   & 5.9 \\
\hline
\hline
\end{tabular}
\end{center}
\end{table}

Following the procedures used in Ref.~\cite{LHCb-PAPER-2014-047}, the relative uncertainty on the luminosity 
is determined to be 1.7\% for the 7\tev data and 3.9\% for the 13\tev data sample.
The relative uncertainty on ${\BR} \left( \Bpm\rightarrow \jpsi \Kpm\right)$
is 3.9\%{~\cite{Abe:2002rc,Aubert:2004rz}},
while the uncertainty on ${\BR} \left( \jpsi \rightarrow \mumu\right)$~\cite{PDG2016} is negligible. 

The variation of the efficiency within a bin induces an uncertainty if the kinematic 
distributions of the simulated samples do not  
match those of the data. This uncertainty, which is important close to the edges 
of the fiducial region, is estimated
by increasing or decreasing the bin width in \pt and $y$ by a factor of two.
The largest variation of the integrated cross-section measurement is
taken as a systematic uncertainty on the production cross-sections.

The systematic uncertainty associated with the invariant mass fits
is obtained by performing fits using alternative choices for the signal and background
functions. The signal PDF is replaced by a {\sc Hypatia} function~\cite{Santos:2013gra},
while the combinatorial background model is modified to be 
either a first-order or second-order polynomial function.
The largest resulting variation of the cross-section measurement is taken as
a systematic uncertainty. 

The efficiencies of the tracking, PID and trigger are estimated using control samples,
and systematic uncertainties arise due to the limited sample sizes.
An additional uncertainty arises on the track-reconstruction efficiency due to limited knowledge of 
the material budget of the detector, which induces a 1.1\% uncertainty on the kaon
reconstruction efficiency due to modelling of hadronic interactions with the detector
material. There is an additional systematic uncertainty 
on the tracking efficiency from the method~\cite{LHCb-DP-2013-002}, which amounts to ${0.4\%~(0.8\%)}$ at ${7\tev~(13\tev)}$.
For the uncertainty from the PID efficiency, the binning effect is studied by 
enlarging or decreasing the number of bins by a factor of two 
in the calculation of the PID efficiency,
and taking the largest deviation from the default as the uncertainty.
An additional uncertainty on the trigger efficiency is determined
by testing the procedure in simulation and taking the deviation as the systematic uncertainty.

The GEC efficiency is obtained from data, and the inefficiency of the global event cuts
is taken as a systematic uncertainty.
The uncertainty on the offline event selection efficiencies are estimated from data and simulation
by varying the selection criteria,
comparing the ratios of the selection and reconstruction efficiencies between data and simulation,
and taking the largest deviation as the systematic uncertainty.

A weighting procedure is applied to the simulation sample to
correct for discrepancies between data and simulation in the track multiplicities.
The weighting factors of the simulated events are varied within their statistical uncertainties and
the largest deviation of the measured cross-section is taken as a systematic uncertainty.

\section{Results}
The measured \Bpm~meson production cross-sections for the 7\tev and 13\tev data
in the range ${0<\pt < 40\gevc}$ and ${2.0<y<4.5}$ are
       \begin{alignat*}{3}
        \sigma(pp\to\Bpm X, \sqrt{s} &=   &\text{7}\tev&) & &=  43.0 \pm 0.2 \pm 2.5 \pm 1.7\mub, \\
        \sigma(pp\to\Bpm X, \sqrt{s} &=   \text{1}&\text{3}\tev&) & &=  86.6 \pm 0.5 \pm 5.4 \pm 3.4\mub,
        \end{alignat*}
 where the first uncertainties are statistical, the second are systematic, and the third are due to the limited knowledge of the \bujpsiks~branching fraction. 

\begin{figure}
\centering
 \includegraphics[width=\bplusoneplotwidth]{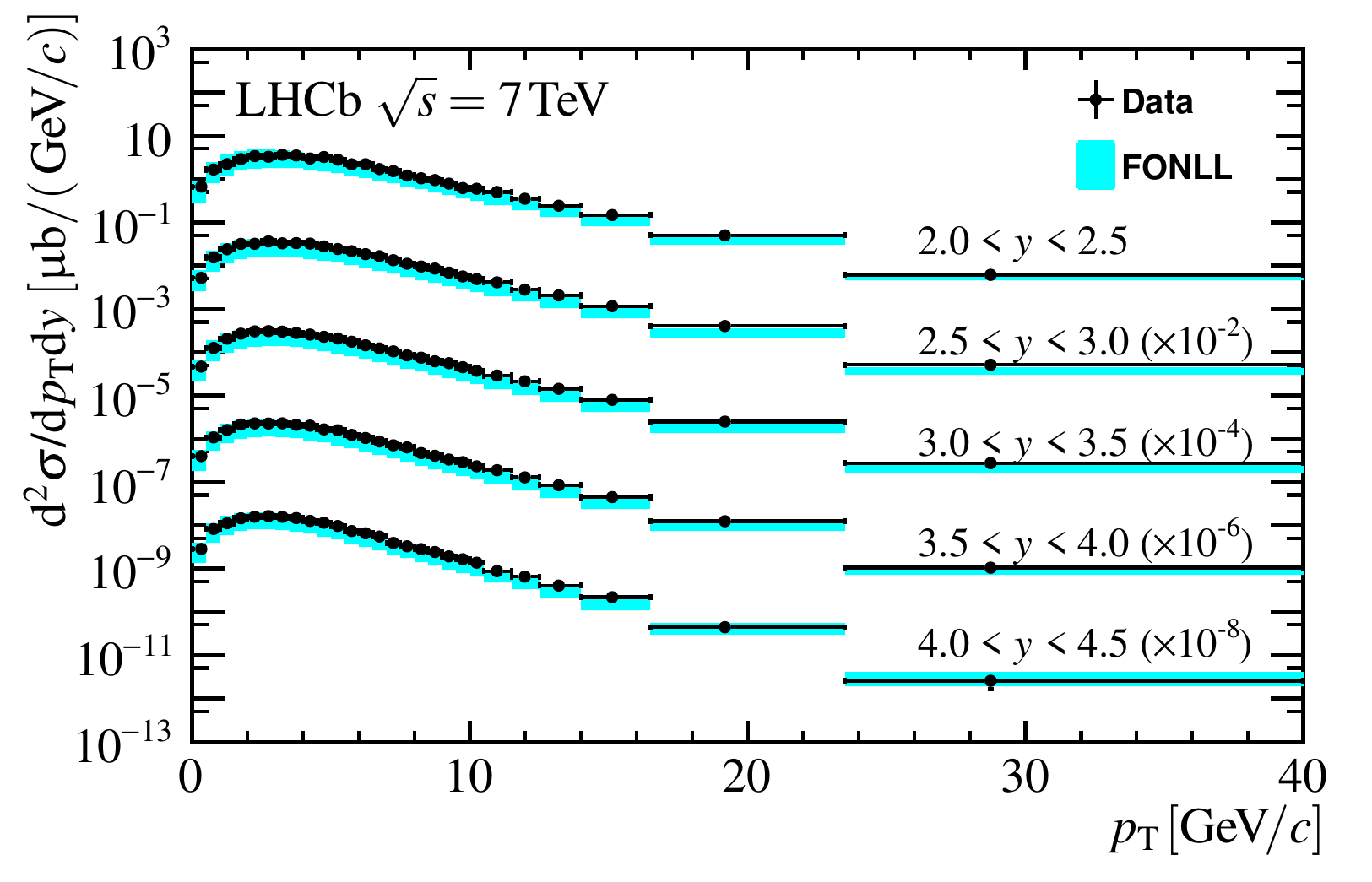}
 \includegraphics[width=\bplusoneplotwidth]{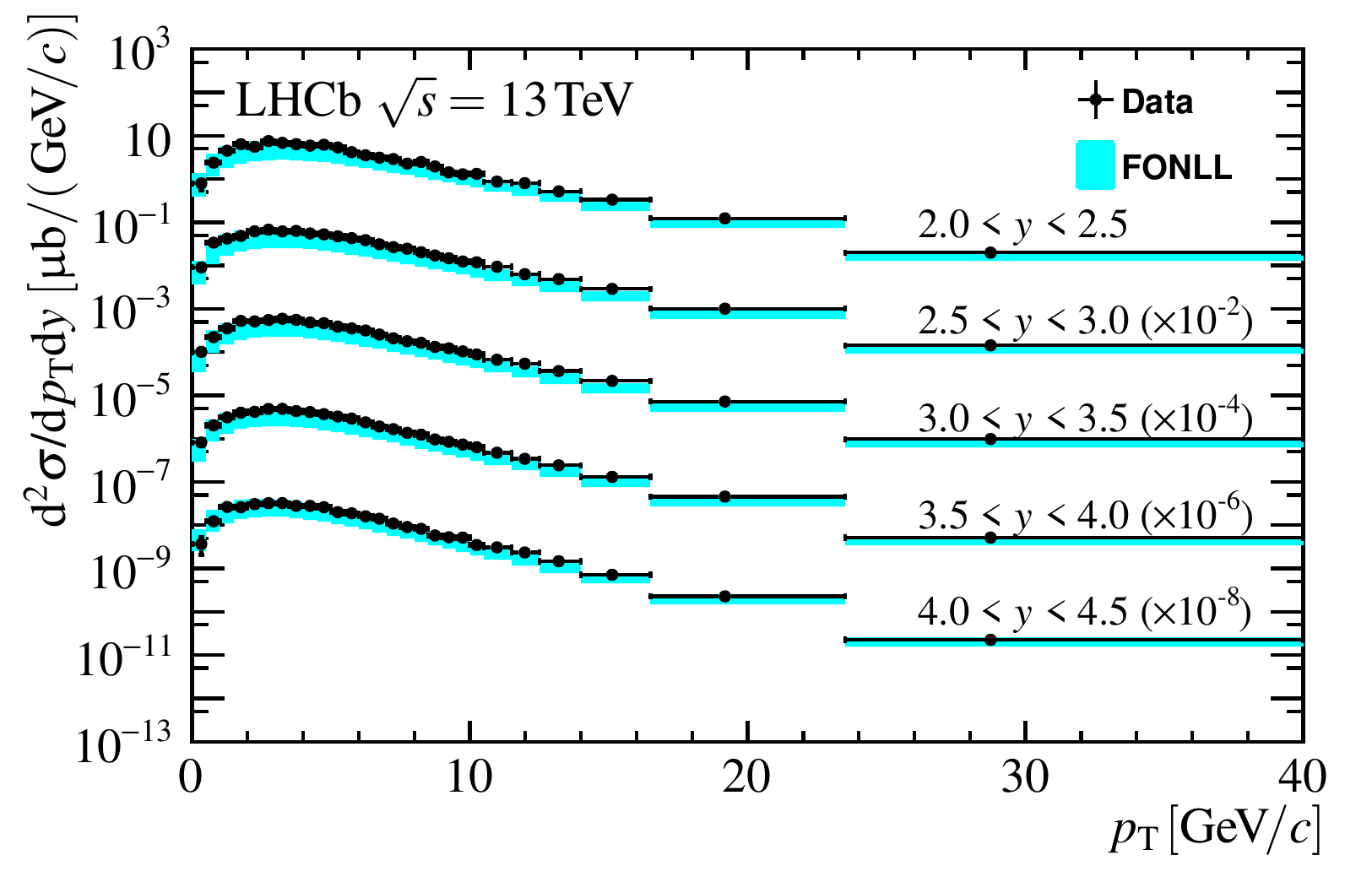}
\caption{Measured \Bpm double-differential production cross-sections at (top) $7\protect\tev$ and (bottom) $13\protect\tev$ as a function of \pt and $y$. The black points represent the measured values,
and the cyan bands are the FONLL predictions~\cite{Cacciari:2015fta}.
Each set of measurements and predictions in a given rapidity bin
is offset by a multiplicative factor $10^{-m}$, where the offset factor is shown after the rapidity range.
The error bars include both the statistical and systematic uncertainties.}
\label{fig:2dxsec7}
\end{figure}

The measured double-differential cross-sections at 7\tev and 13\tev
as functions of $\pt$ and $y$ are shown in Fig.~\ref{fig:2dxsec7},
where the measurements are compared with theoretical predictions
based on FONLL calculations~\cite{Cacciari:2015fta}.
The predictions depend on assumptions on the $b$-quark mass,
the renormalization and factorization scales, and parton distribution functions (PDFs).
The $b$-quark production cross-section calculated with the FONLL approach uses
a $b$-quark mass of 4.75\gevcc,
the renormalization and factorization scales, $\mu_R$, $\mu_F$, 
set to $\mu_R=\mu_F=\mu_0=\sqrt{m^2_Q+\pt^2}$, where $m_Q$ and $\pt$ are
the mass and transverse momenta of the $b$ quark, and the CTEQ6.6~\cite{Nadolsky:2008zw} PDFs.
Varying the $b$-quark mass by $\pm0.25\gevcc$, 
the $\mu_R/\mu_F$ ratio by a factor of two, or using
different PDFs introduces an uncertainty in the predicted $b$-quark production 
cross-section of up to 50\% at low \pt~\cite{Cacciari:2015fta}.

\begin{figure}
 \centering
 \includegraphics[width=\bplusplotwidth]{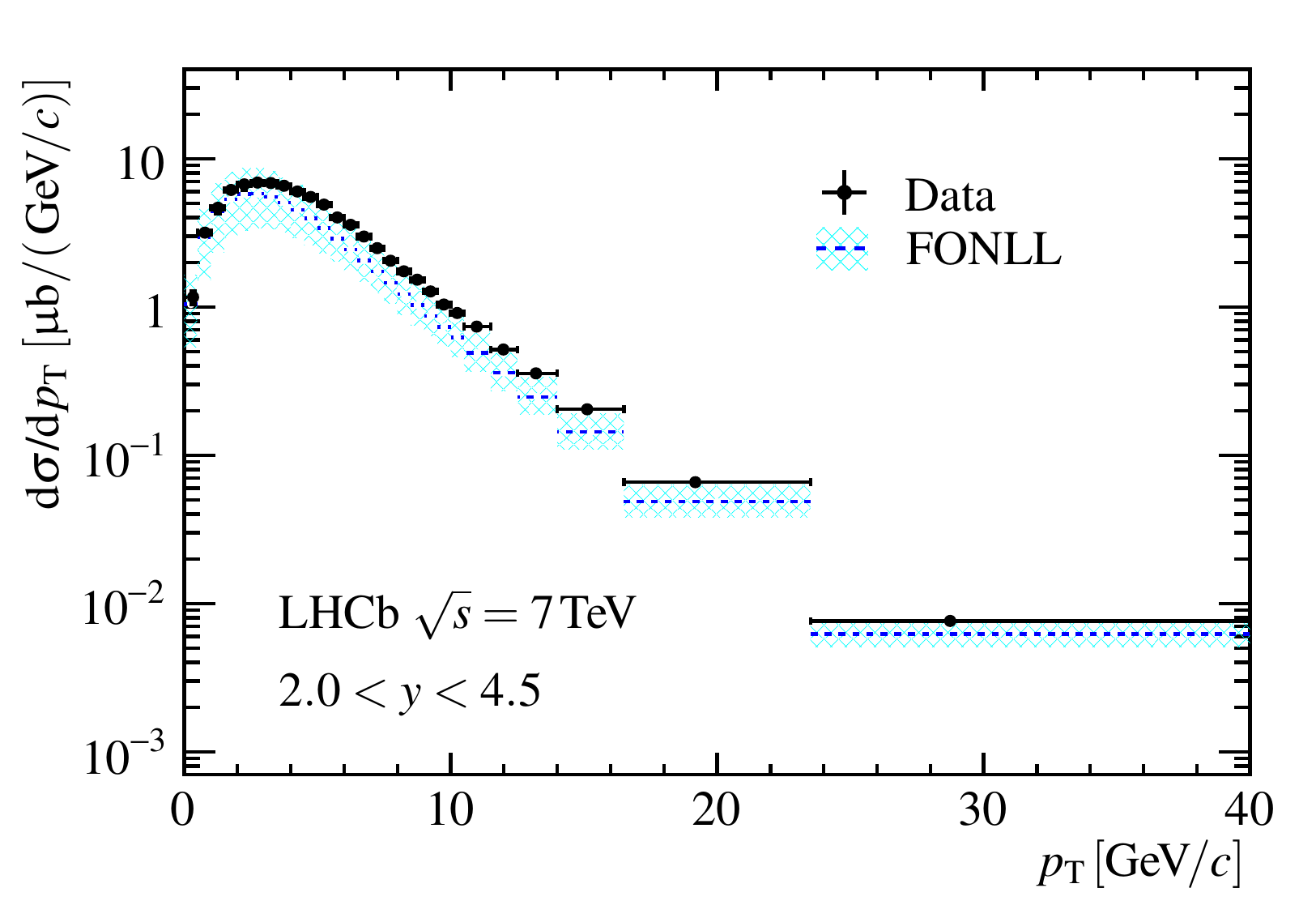}
 \includegraphics[width=\bplusplotwidth]{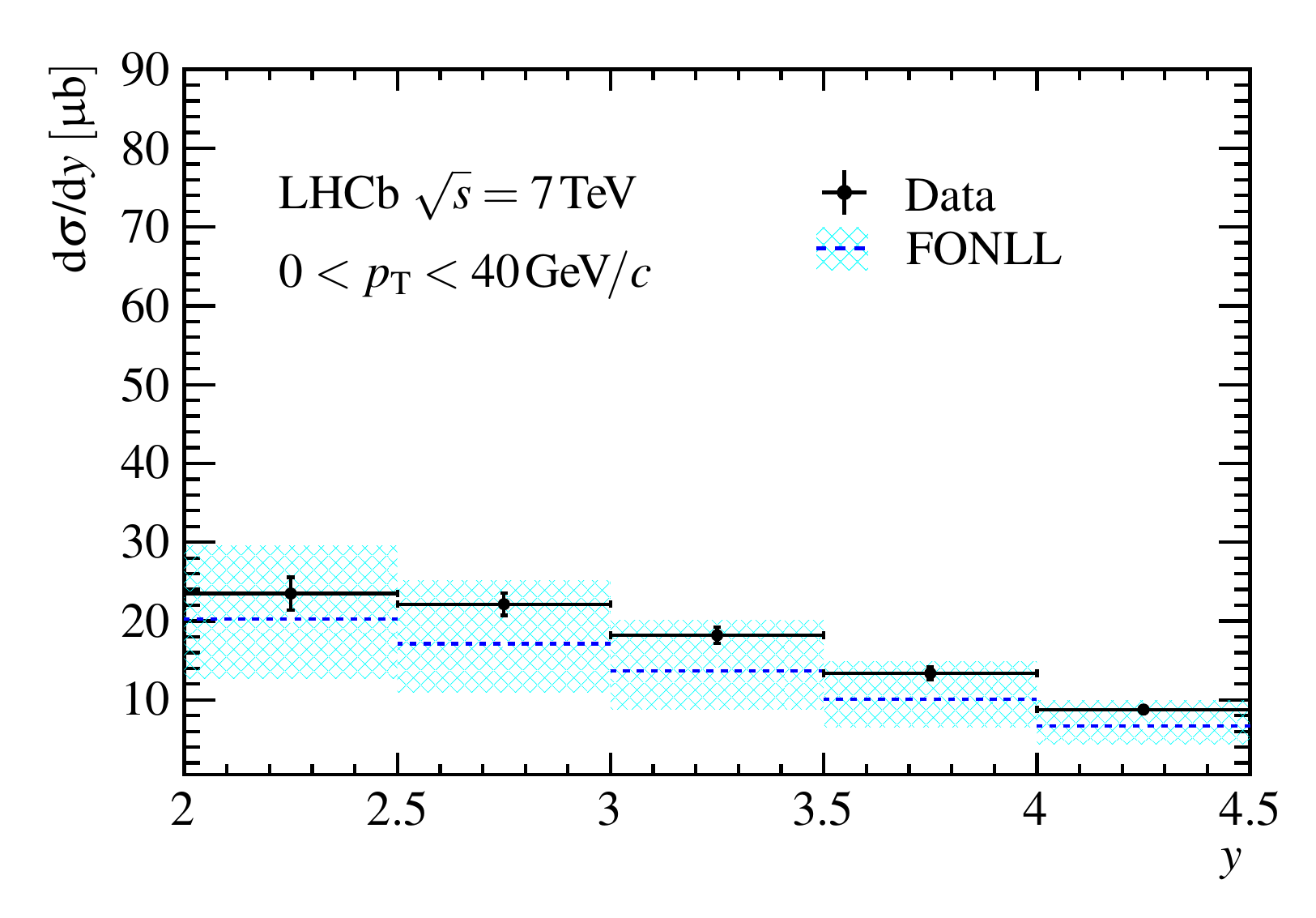}
 \caption{Measured \Bpm differential cross-section at $7\protect\tev$ as a function of (left) \pt or (right) $y$.
The black points represent the measured values,
the blue-dashed line and cyan band represent the central values and uncertainties of the FONLL prediction~\cite{Cacciari:2015fta}.
The error bars include both the statistical and systematic uncertainties.}
\label{fig:1dxsec7}
\end{figure}

\begin{figure}
 \centering
 \includegraphics[width=\bplusplotwidth]{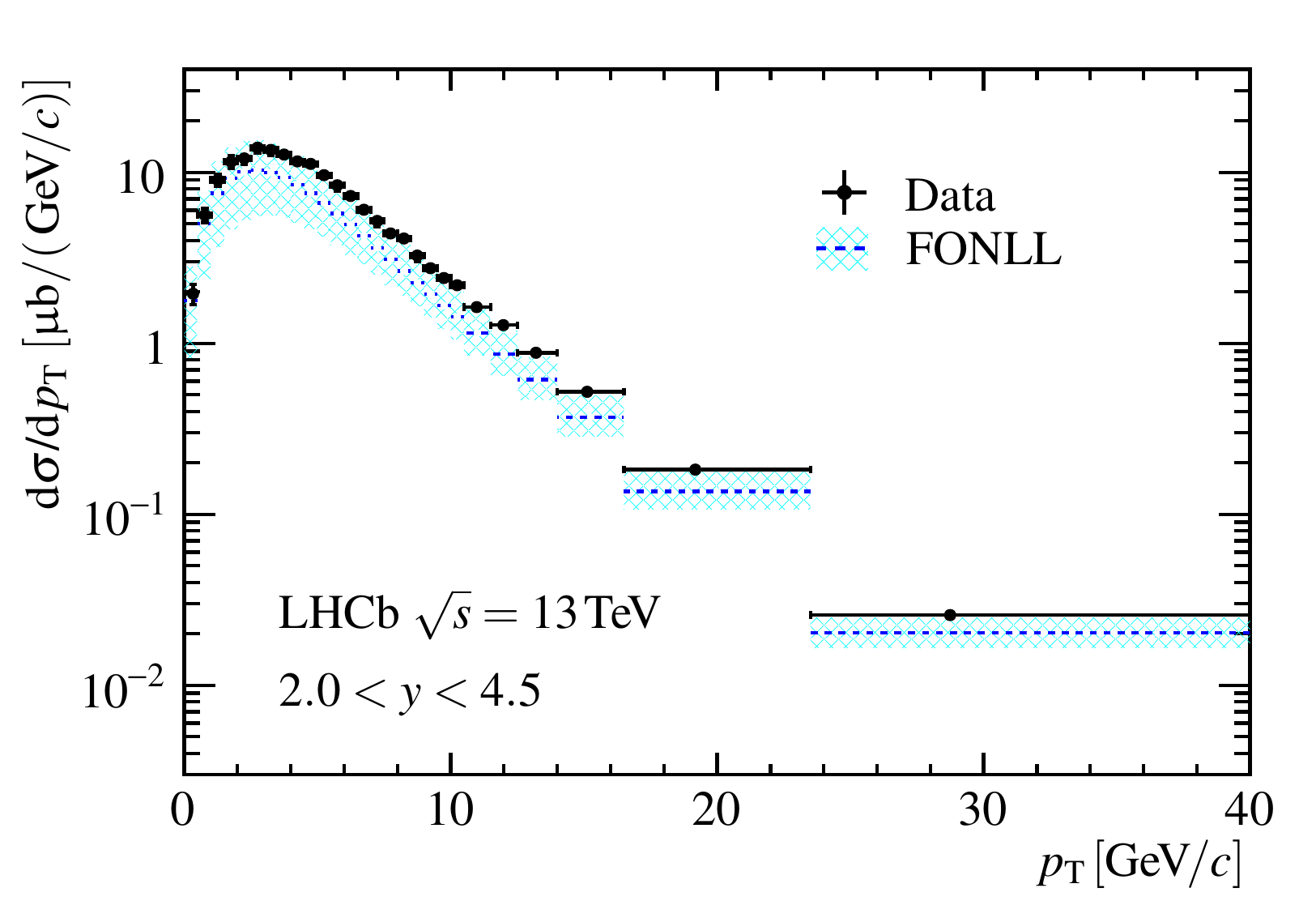}
 \includegraphics[width=\bplusplotwidth]{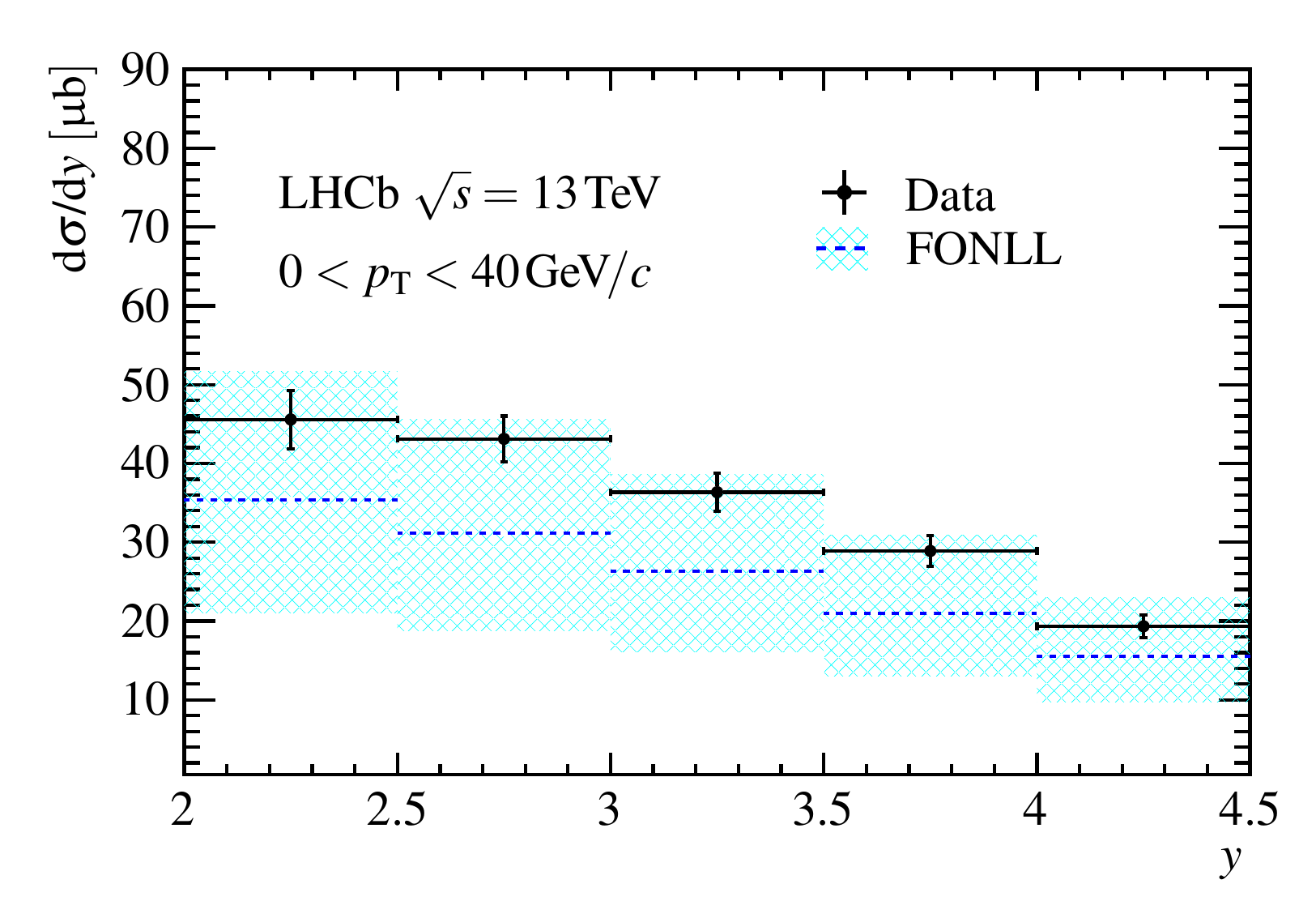}
 \caption{Measured \Bpm differential cross-section at $13\protect\tev$ as a function of (left) \pt or (right) $y$.
The black points represent the measured values,
the blue-dashed line and cyan band represent the central values and uncertainties of the FONLL prediction~\cite{Cacciari:2015fta}. 
The error bars include both the statistical and systematic uncertainties.}
\label{fig:1dxsec}
\end{figure}

The corresponding single-differential cross-sections
are shown in Figs.~\ref{fig:1dxsec7} and~\ref{fig:1dxsec}.
The single-differential cross-sections and the production cross-section in the above 
range of \pt and $y$ are calculated from the measured double-differential cross-sections.
All results are in agreement with the FONLL predictions.
The results are tabulated in Appendix~\ref{sec:Supplementary-App}.

\begin{figure}
 \centering
 \includegraphics[width=\bplusplotwidth]{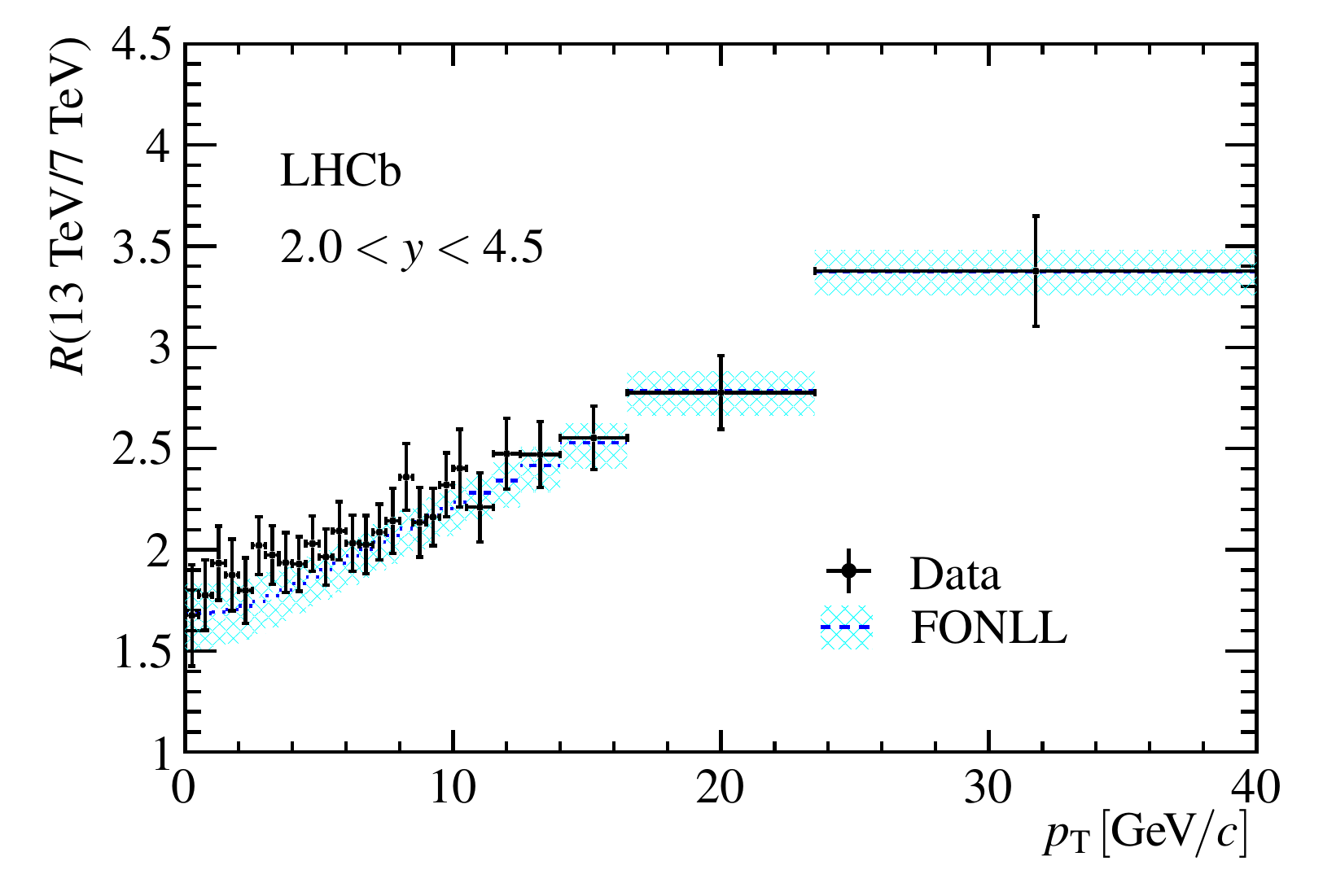}
 \includegraphics[width=\bplusplotwidth]{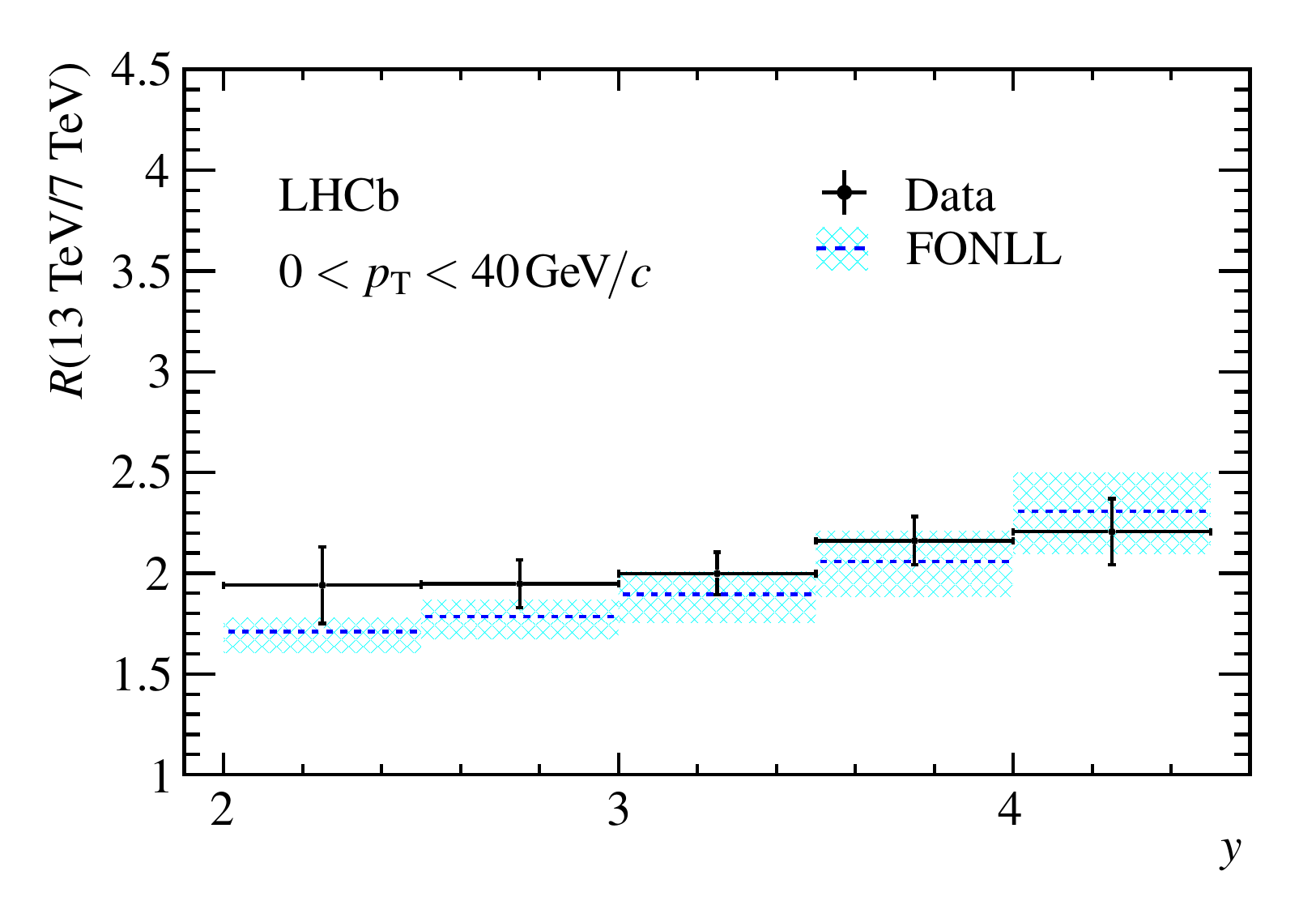}
 \caption{Ratio of the \Bpm cross-section at $13\protect\tev$ to that at $7\protect\tev$
 as a function of (left) \pt or (right) $y$. The black points represent the measured values, 
 the blue-dashed line and cyan band represent the central values and uncertainties of the FONLL prediction~\cite{Cacciari:2015fta}.}
 \label{fig:comp_137_1d}
\end{figure}

The ratio of the cross-section at 13\tev to that at 7\tev, $R(13\tev/7\tev)$,
 is determined to be
 \begin{eqnarray*}
 R(13\tev/7\tev) = \ratiores\,.
 \end{eqnarray*}
In the ratio calculation, the systematic uncertainties on the luminosities at 13 and 7\tev
 are taken to be 50\% correlated, as in Ref.~\cite{LHCb-PAPER-2014-047};
the systematic uncertainties associated with the branching fractions, mass fits, event selection and binning
are assumed to be completely correlated;
and all other uncertainties are considered to be uncorrelated.
The systematic uncertainty on $R$ is summarised in Table~\ref{tab:uncer_tot}.
In Fig.~\ref{fig:comp_137_1d}, the ratio of
the cross-section at 13\tev to that at 7\tev as a function of \pt or $y$ is
compared with the FONLL predictions. 
The measured results agree with the FONLL
predictions in both the shape and the scale.

\section{Summary}
In summary, the double-differential production cross-sections of \Bpm mesons are measured as functions of the transverse momentum and rapidity, using $pp$ collision data collected with the LHCb detector at the Large Hadron Collider.
The integrated luminosities of the data samples are $1.0\invfb$ and $0.3\invfb$
at the centre-of-mass energies of $7\tev$ and $13\tev$, respectively.
The measurements are performed in the transverse momentum range $0<\pt<40\gevc$ and the rapidity range $2.0<y<4.5$. 
The $7\tev$ results are consistent with previously
published results{~\cite{LHCb-PAPER-2011-043,LHCb-PAPER-2013-004}},
with improved precision in the low $y$ region. This measurement supersedes previous results.
The ratio of the production cross-section at $13\tev$ to that at $7\tev$ is also measured.
All results are in agreement with theoretical calculations based on the FONLL approach.
\input{acknowledgements}

\input{appendix}

\addcontentsline{toc}{section}{References}
\setboolean{inbibliography}{true}
\bibliographystyle{LHCb}
\bibliography{Bplus,main,LHCb-PAPER,LHCb-CONF,LHCb-DP,LHCb-TDR}

\newpage

\newpage
\input{LHCb_Authorship_flat_10-Aug-2017}
\end{document}

%% file: preamble.tex
\usepackage[top=1in, bottom=1.25in, left=1in, right=1in]{geometry}

\usepackage{microtype}
\usepackage{lineno}  
\usepackage{xspace} 
\usepackage{caption} 

\usepackage{graphicx}  
\usepackage{color}
\usepackage{colortbl}
\graphicspath{{./figs/}} 

\usepackage{amsmath} 
\usepackage{amssymb}
\usepackage{amsfonts}
\usepackage{upgreek} 

\newcommand*\patchAmsMathEnvironmentForLineno[1]{%
\expandafter\let\csname old#1\expandafter\endcsname\csname #1\endcsname
\expandafter\let\csname oldend#1\expandafter\endcsname\csname
end#1\endcsname
 \renewenvironment{#1}%
   {\linenomath\csname old#1\endcsname}%
   {\csname oldend#1\endcsname\endlinenomath}%
}
\newcommand*\patchBothAmsMathEnvironmentsForLineno[1]{%
  \patchAmsMathEnvironmentForLineno{#1}%
  \patchAmsMathEnvironmentForLineno{#1*}%
}
\AtBeginDocument{%
\patchBothAmsMathEnvironmentsForLineno{equation}%
\patchBothAmsMathEnvironmentsForLineno{align}%
\patchBothAmsMathEnvironmentsForLineno{flalign}%
\patchBothAmsMathEnvironmentsForLineno{alignat}%
\patchBothAmsMathEnvironmentsForLineno{gather}%
\patchBothAmsMathEnvironmentsForLineno{multline}%
\patchBothAmsMathEnvironmentsForLineno{eqnarray}%
}

\usepackage{hyperxmp}

\usepackage[pdftex,
            pdfauthor={\paperauthors},
            pdftitle={\paperasciititle},
            pdfkeywords={\paperkeywords},
            pdfcopyright={Copyright (C) \papercopyright},
            pdflicenseurl={\paperlicenceurl}]{hyperref}

\usepackage[all]{hypcap} 

\input{lhcb-symbols-def} 

\usepackage{cite} 
\usepackage{mciteplus}

%% file: lhcb-symbols-def.tex
\usepackage{xspace} 
\usepackage{upgreek}

\def\lhcb {\mbox{LHCb}\xspace}
\def\atlas  {\mbox{ATLAS}\xspace}
\def\cms    {\mbox{CMS}\xspace}

\def\babar  {\mbox{BaBar}\xspace}
\def\belle  {\mbox{Belle}\xspace}

\def\MagUp {\mbox{\em Mag\kern -0.05em Up}\xspace}

\ifthenelse{\boolean{uprightparticles}}%
{

 \def\Pmu         {\ensuremath{\upmu}\xspace}

 \def\Ppi         {\ensuremath{\uppi}\xspace}

 \def\Ppsi        {\ensuremath{\uppsi}\xspace}

 \def\PDelta      {\ensuremath{\Delta}\xspace}                 
 \def\PXi      {\ensuremath{\Xi}\xspace}                 
 \def\PLambda      {\ensuremath{\Lambda}\xspace}                 
 \def\PSigma      {\ensuremath{\Sigma}\xspace}                 
 \def\POmega      {\ensuremath{\Omega}\xspace}                 
 \def\PUpsilon      {\ensuremath{\Upsilon}\xspace}                 
 
 \def\PB      {\ensuremath{\mathrm{B}}\xspace}                 
                  
 \def\PD      {\ensuremath{\mathrm{D}}\xspace}

 \def\PJ      {\ensuremath{\mathrm{J}}\xspace}                 
 \def\PK      {\ensuremath{\mathrm{K}}\xspace}

 \def\Pb      {\ensuremath{\mathrm{b}}\xspace}                 
 \def\Pc      {\ensuremath{\mathrm{c}}\xspace}

 \def\Pi      {\ensuremath{\mathrm{i}}\xspace}

}
{

 \def\Pmu         {\ensuremath{\mu}\xspace}

 \def\Ppi         {\ensuremath{\pi}\xspace}

 \def\Ppsi        {\ensuremath{\psi}\xspace}                 
                  
 \mathchardef\PDelta="7101
 \mathchardef\PXi="7104
 \mathchardef\PLambda="7103
 \mathchardef\PSigma="7106
 \mathchardef\POmega="710A
 \mathchardef\PUpsilon="7107
                  
 \def\PB      {\ensuremath{B}\xspace}                 
                  
 \def\PD      {\ensuremath{D}\xspace}

 \def\PJ      {\ensuremath{J}\xspace}                 
 \def\PK      {\ensuremath{K}\xspace}

 \def\Pb      {\ensuremath{b}\xspace}                 
 \def\Pc      {\ensuremath{c}\xspace}

 \def\Pi      {\ensuremath{i}\xspace}

}

\makeatletter
\ifcase \@ptsize \relax
  \newcommand{\miniscule}{\@setfontsize\miniscule{4}{5}}
\or
  \newcommand{\miniscule}{\@setfontsize\miniscule{5}{6}}
\or
  \newcommand{\miniscule}{\@setfontsize\miniscule{5}{6}}
\fi
\makeatother

\DeclareRobustCommand{\optbar}[1]{\shortstack{{\miniscule (\rule[.5ex]{1.25em}{.18mm})}
  \\ [-.7ex] $#1$}}

\def\mup        {{\ensuremath{\Pmu^+}}\xspace}
\def\mun        {{\ensuremath{\Pmu^-}}\xspace} 
\def\mumu       {{\ensuremath{\Pmu^+\Pmu^-}}\xspace}

\def\cquark    {{\ensuremath{\Pc}}\xspace}

\def\bquark    {{\ensuremath{\Pb}}\xspace}

\def\pion   {{\ensuremath{\Ppi}}\xspace}

\def\pipm   {{\ensuremath{\pion^\pm}}\xspace}

\def\kaon    {{\ensuremath{\PK}}\xspace}
  \def\Kbar    {{\kern 0.2em\overline{\kern -0.2em \PK}{}}\xspace}

\def\KorKbar    {\kern 0.18em\optbar{\kern -0.18em K}{}\xspace}

\def\Kpm     {{\ensuremath{\kaon^\pm}}\xspace}

  \def\Dbar    {{\kern 0.2em\overline{\kern -0.2em \PD}{}}\xspace}

\def\DorDbar    {\kern 0.18em\optbar{\kern -0.18em D}{}\xspace}

\def\B       {{\ensuremath{\PB}}\xspace}
\def\Bbar    {{\ensuremath{\kern 0.18em\overline{\kern -0.18em \PB}{}}}\xspace}

\def\BorBbar    {\kern 0.18em\optbar{\kern -0.18em B}{}\xspace}

\def\Bpm     {{\ensuremath{\B^\pm}}\xspace}

\def\jpsi     {{\ensuremath{{\PJ\mskip -3mu/\mskip -2mu\Ppsi\mskip 2mu}}}\xspace}

  \def\Y#1S{\ensuremath{\PUpsilon{(#1S)}}\xspace}

\def\Lbar        {{\ensuremath{\kern 0.1em\overline{\kern -0.1em\PLambda}}}\xspace}
\def\LorLbar    {\kern 0.18em\optbar{\kern -0.18em \PLambda}{}\xspace}

\def\BF         {{\ensuremath{\mathcal{B}}}\xspace}

\def\BR         {\BF}

\def\to                 {\ensuremath{\rightarrow}\xspace}

\def\eps   {{\ensuremath{\varepsilon}}\xspace}

\def\AT#1     {\ensuremath{A_{\mathrm{T}}^{#1}}\xspace}           

\def\C#1      {\ensuremath{\mathcal{C}_{#1}}\xspace}                       
\def\Cp#1     {\ensuremath{\mathcal{C}_{#1}^{'}}\xspace}                    
\def\Ceff#1   {\ensuremath{\mathcal{C}_{#1}^{\mathrm{(eff)}}}\xspace}        
\def\Cpeff#1  {\ensuremath{\mathcal{C}_{#1}^{'\mathrm{(eff)}}}\xspace}       
\def\Ope#1    {\ensuremath{\mathcal{O}_{#1}}\xspace}                       
\def\Opep#1   {\ensuremath{\mathcal{O}_{#1}^{'}}\xspace}                    

\newcommand{\tev}{\ifthenelse{\boolean{inbibliography}}{\ensuremath{~T\kern -0.05em eV}}{\ensuremath{\mathrm{\,Te\kern -0.1em V}}}\xspace}
\newcommand{\gev}{\ensuremath{\mathrm{\,Ge\kern -0.1em V}}\xspace}
\newcommand{\mev}{\ensuremath{\mathrm{\,Me\kern -0.1em V}}\xspace}
\newcommand{\kev}{\ensuremath{\mathrm{\,ke\kern -0.1em V}}\xspace}
\newcommand{\ev}{\ensuremath{\mathrm{\,e\kern -0.1em V}}\xspace}
\newcommand{\gevc}{\ensuremath{{\mathrm{\,Ge\kern -0.1em V\!/}c}}\xspace}
\newcommand{\mevc}{\ensuremath{{\mathrm{\,Me\kern -0.1em V\!/}c}}\xspace}
\newcommand{\gevcc}{\ensuremath{{\mathrm{\,Ge\kern -0.1em V\!/}c^2}}\xspace}
\newcommand{\gevgevcccc}{\ensuremath{{\mathrm{\,Ge\kern -0.1em V^2\!/}c^4}}\xspace}
\newcommand{\mevcc}{\ensuremath{{\mathrm{\,Me\kern -0.1em V\!/}c^2}}\xspace}

\def\mum  {\ensuremath{{\,\upmu\mathrm{m}}}\xspace}

\def\mub{\ensuremath{{\mathrm{ \,\upmu b}}}\xspace}
\def\nb {\ensuremath{\mathrm{ \,nb}}\xspace}

\def\invpb {\ensuremath{\mbox{\,pb}^{-1}}\xspace}

\def\invfb   {\ensuremath{\mbox{\,fb}^{-1}}\xspace}

\def\ps   {\ensuremath{{\mathrm{ \,ps}}}\xspace}

\newcommand{\stat}{\ensuremath{\mathrm{\,(stat)}}\xspace}
\newcommand{\syst}{\ensuremath{\mathrm{\,(syst)}}\xspace}

\def\deriv {\ensuremath{\mathrm{d}}}

\def\gsim{{~\raise.15em\hbox{$>$}\kern-.85em
          \lower.35em\hbox{$\sim$}~}\xspace}
\def\lsim{{~\raise.15em\hbox{$<$}\kern-.85em
          \lower.35em\hbox{$\sim$}~}\xspace}

\def\ptot       {\mbox{$p$}\xspace}
\def\pt         {\mbox{$p_{\mathrm{ T}}$}\xspace}

\newcommand{\lum} {\ensuremath{\mathcal{L}}\xspace}

\def\evtgen     {\mbox{\textsc{EvtGen}}\xspace}

\def\geant      {\mbox{\textsc{Geant4}}\xspace}

\def\photos     {\mbox{\textsc{Photos}}\xspace}

\def\tell1  {TELL1\xspace}
\def\ukl1   {UKL1\xspace}



%% file: title-LHCb-PAPER.tex

\begin{titlepage}
\pagenumbering{roman}

\vspace*{-1.5cm}
\centerline{\large EUROPEAN ORGANIZATION FOR NUCLEAR RESEARCH (CERN)}
\vspace*{1.5cm}
\noindent
\begin{tabular*}{\linewidth}{lc@{\extracolsep{\fill}}r@{\extracolsep{0pt}}}
\ifthenelse{\boolean{pdflatex}}
{\vspace*{-2.7cm}\mbox{\!\!\!\includegraphics[width=.14\textwidth]{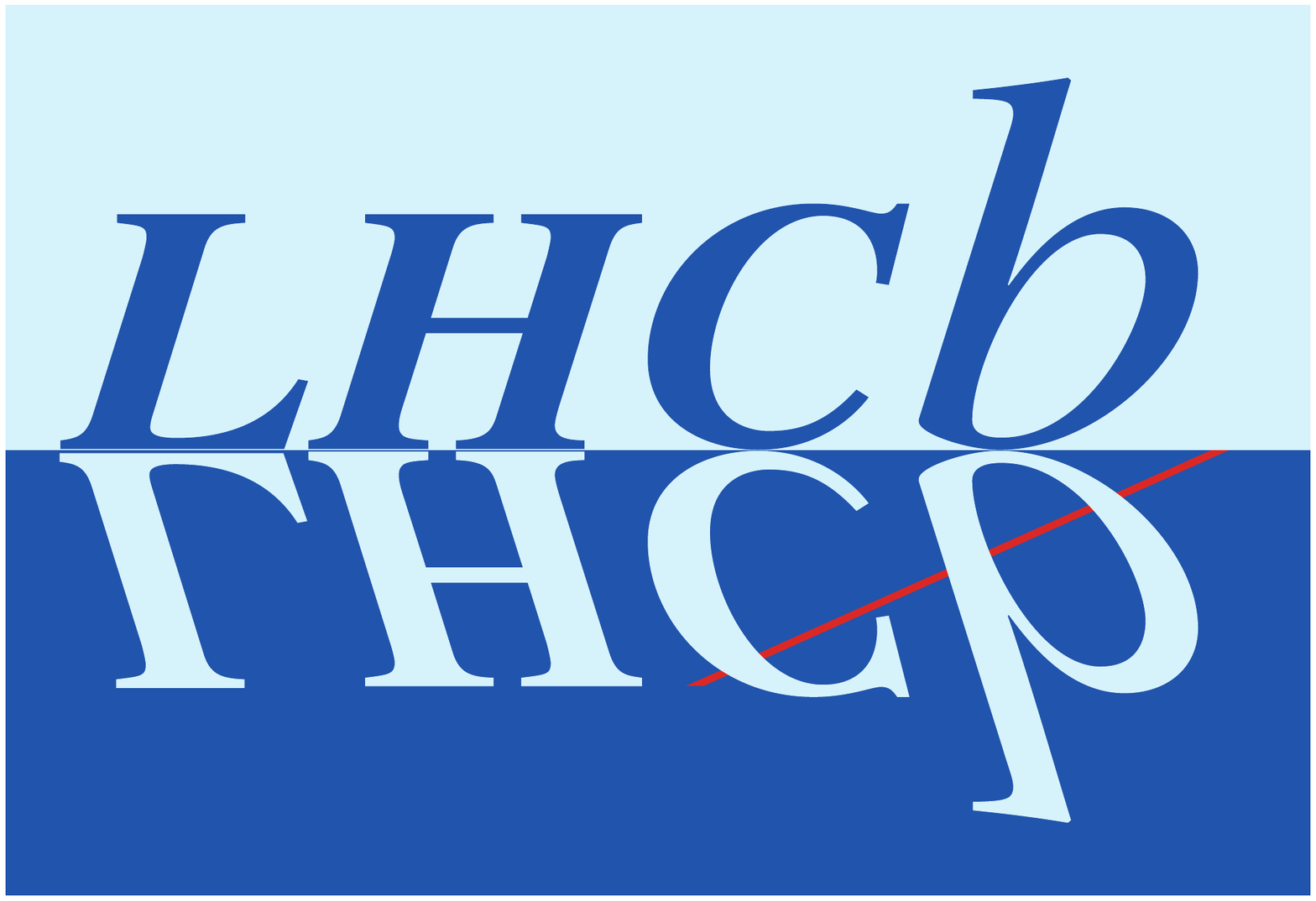}} & &}%
{\vspace*{-1.2cm}\mbox{\!\!\!\includegraphics[width=.12\textwidth]{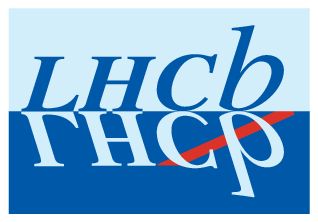}} & &}%
\\
 & & CERN-EP-2017-254 \\  
 & & LHCb-PAPER-2017-037 \\  
 & & 07 December 2017 \\ 
\end{tabular*}

\vspace*{2.0cm}

{\normalfont\bfseries\boldmath\huge
\begin{center}
  \papertitle 
\end{center}
}

\vspace*{0.5cm}

\begin{center}
\paperauthors\footnote{Authors are listed at the end of this paper.}
\end{center}

\begin{abstract}
  \noindent
        The production of \Bpm~mesons is studied in $pp$ collisions
        at {centre-of-mass} energies of 7 and 13\tev,
        using \bujpsiks~decays and data samples corresponding to 1.0\invfb and 0.3\invfb, respectively. 
        The production cross-sections 
        summed over both charges and integrated over 
        the transverse momentum range ${0<\pt< 40\gevc}$
        and the rapidity range ${2.0<y<4.5}$ are measured to be
       \begin{alignat*}{3}
        \sigma(pp\to\Bpm X, \sqrt{s} &=   &\text{7}\tev&) & &=  43.0 \pm 0.2 \pm 2.5 \pm 1.7\mub, \\
        \sigma(pp\to\Bpm X, \sqrt{s} &=   \text{1}&\text{3}\tev&) & &=  86.6 \pm 0.5 \pm 5.4 \pm 3.4\mub,
        \end{alignat*}
        where the first uncertainties are statistical, the second are systematic, 
        and the third are due to the limited knowledge of the \bujpsiks~branching fraction. 
        The ratio of the cross-section at 13\tev to that at 7\tev is determined to be
        \ratiores.
        Differential cross-sections are also reported as functions of \pt and $y$.
        All results are in agreement with theoretical calculations
        based on the state-of-art fixed next-to-leading order quantum chromodynamics.
  \end{abstract}

\vspace*{0.5cm}

\begin{center}
  Published in JHEP 12 (2017) 026
\end{center}

\vspace{\fill}

{\footnotesize
\centerline{\copyright~\papercopyright, licence \href{\paperlicenceurl}{\paperlicence}.}}
\vspace*{2mm}

\end{titlepage}


\newpage
\setcounter{page}{2}
\mbox{~}

\cleardoublepage

%% file: acknowledgements.tex
\section*{Acknowledgements}
%
%
\noindent We express our gratitude to our colleagues in the CERN
accelerator departments for the excellent performance of the LHC. We
thank the technical and administrative staff at the LHCb
institutes. We acknowledge support from CERN and from the national
agencies: CAPES, CNPq, FAPERJ and FINEP (Brazil); MOST and NSFC
(China); CNRS/IN2P3 (France); BMBF, DFG and MPG (Germany); INFN
(Italy); NWO (The Netherlands); MNiSW and NCN (Poland); MEN/IFA
(Romania); MinES and FASO (Russia); MinECo (Spain); SNSF and SER
(Switzerland); NASU (Ukraine); STFC (United Kingdom); NSF (USA).  We
acknowledge the computing resources that are provided by CERN, IN2P3
(France), KIT and DESY (Germany), INFN (Italy), SURF (The
Netherlands), PIC (Spain), GridPP (United Kingdom), RRCKI and Yandex
LLC (Russia), CSCS (Switzerland), IFIN-HH (Romania), CBPF (Brazil),
PL-GRID (Poland) and OSC (USA). We are indebted to the communities
behind the multiple open-source software packages on which we depend.
Individual groups or members have received support from AvH Foundation
(Germany), EPLANET, Marie Sk\l{}odowska-Curie Actions and ERC
(European Union), ANR, Labex P2IO, ENIGMASS and OCEVU, and R\'{e}gion
Auvergne-Rh\^{o}ne-Alpes (France), RFBR and Yandex LLC (Russia), GVA,
XuntaGal and GENCAT (Spain), Herchel Smith Fund, the Royal Society,
the English-Speaking Union and the Leverhulme Trust (United Kingdom).

%% file: appendix.tex
\appendix

\section{Acceptance efficiency}
\label{app:acc}

\begin{figure}
 \centering
 \includegraphics[width=\bplusplotwidth]{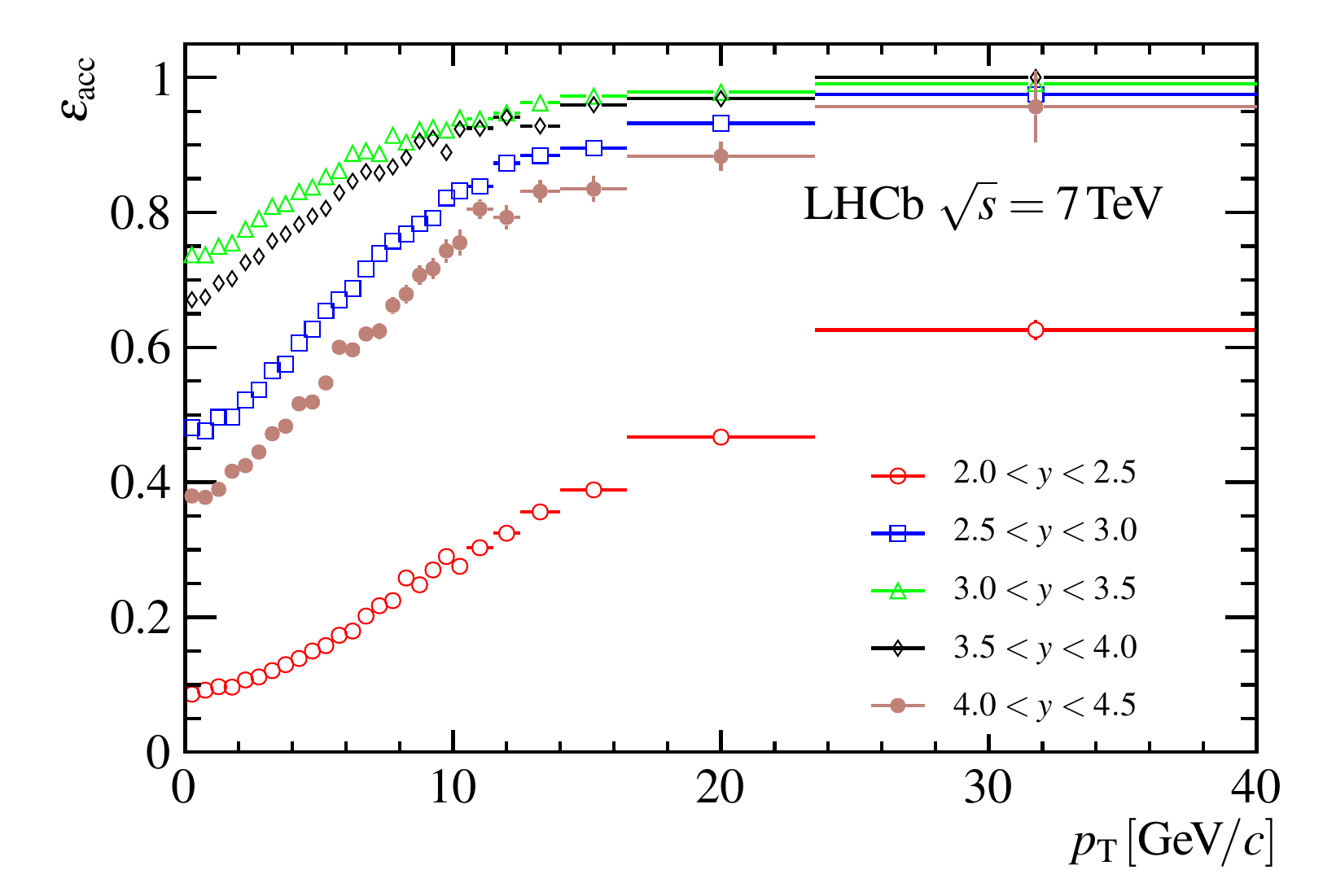}
 \includegraphics[width=\bplusplotwidth]{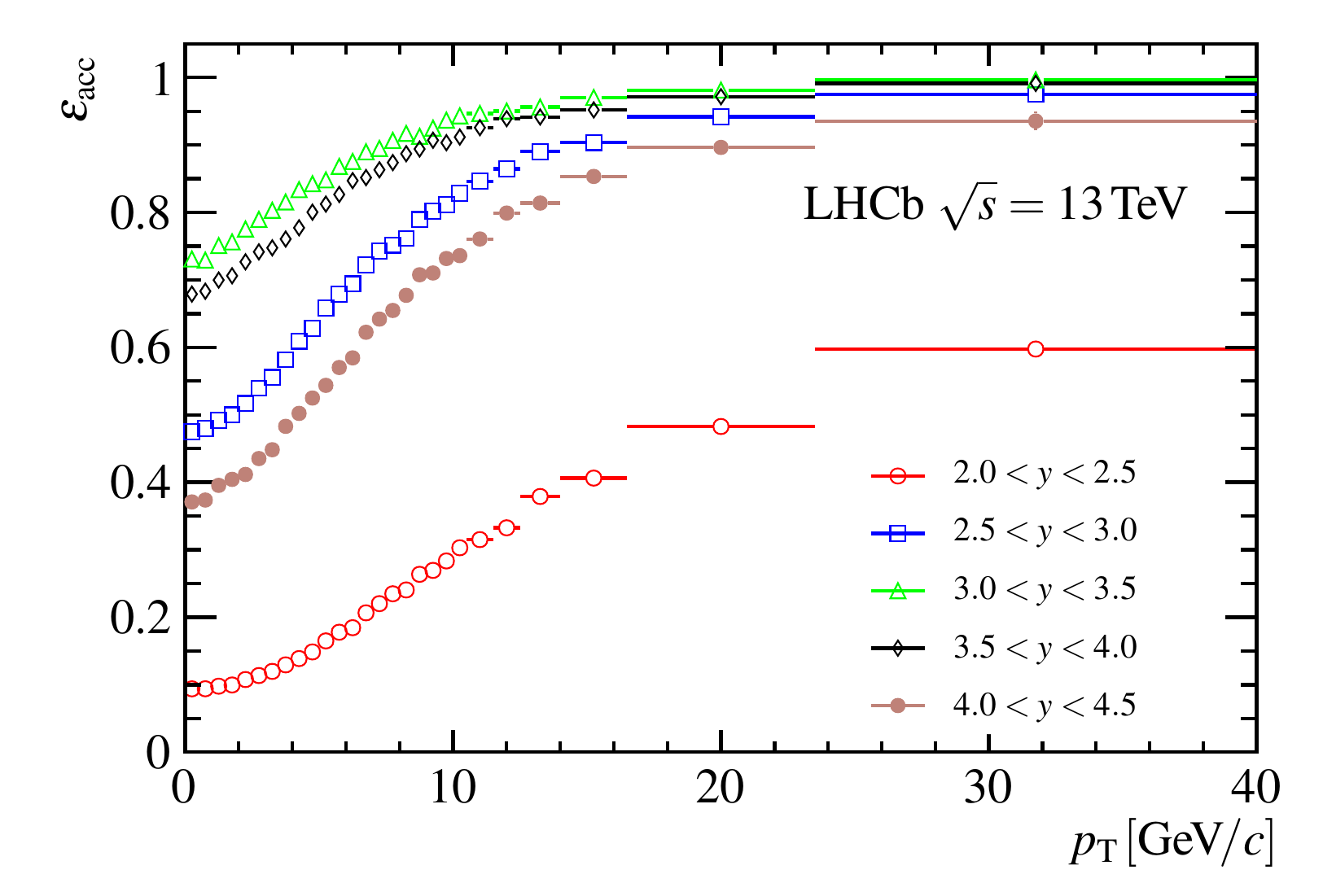}
 \caption{Measured acceptance efficiency of \Bpm events at (left) $7\protect\tev$ and (right) $13\protect\tev$ as a function of \pt in different $y$ regions.}
\label{fig:acc}
\end{figure}

The measured $\varepsilon_{\text{acc}}$ as a function of \pt in different $y$ regions are shown in Fig.~\ref{fig:acc} for 7\tev and 13\tev, respectively.

\section{Total efficiencies}
\label{app:eff}

The measured $\varepsilon_{\text{tot}}$ with its statistical uncertainty as a function of \pt and $y$ are given 
in Tables~\ref{tab:eff_7} and~\ref{tab:eff_13}, for 7\tev and 13\tev, respectively.
The $\varepsilon_{\text{tot}}$ in the low \pt region are smaller than that of the high \pt region, 
due to the lifetime cut effects.
Also with limited detector coverage, $\varepsilon_{\text{tot}}$ in the boundary regions ($2.0<y<2.5$ and $4.0<y<4.5$) 
are smaller than that in central regions.

\begin{table}[htbp]
\begin{center}
\caption{\label{tab:eff_7} Measured $\varepsilon_{\text{tot}}$ of \Bpm events at $7\protect\tev$,
in the bins of \Bpm \pt and $y$. The efficiencies and uncertainties are in percent.}
\begin{tabular}{r@{.}l@{$~-~$}r@{.}lr@{.}l@{$\pm$}r@{.}lr@{.}l@{$\pm$}r@{.}lr@{.}l@{$\pm$}r@{.}lr@{.}l@{$\pm$}r@{.}l r@{.}l@{$\pm$}r@{.}l}
\multicolumn{4}{c}{\pt [\gevc]} & \multicolumn{4}{c}{$2.0<y<2.5$} & \multicolumn{4}{c}{$2.5<y<3.0$} & \multicolumn{4}{c}{$3.0<y<3.5$} & \multicolumn{4}{c}{$3.5<y<4.0$} & \multicolumn{4}{c}{$4.0<y<4.5$}\\ \hline
 0&0 & 0&5  & 0&6 & 0&1  & 5&4 & 0&3  & 10&6 & 0&5  & 9&4 & 0&5  & 4&0 & 0&3 \\ 
 0&5 & 1&0  & 0&7 & 0&0  & 5&3 & 0&2  & 10&3 & 0&4  & 10&0 & 0&5  & 4&2 & 0&3 \\ 
 1&0 & 1&5  & 0&8 & 0&0  & 5&7 & 0&2  & 10&4 & 0&4  & 10&6 & 0&4  & 4&4 & 0&2 \\ 
 1&5 & 2&0  & 0&8 & 0&0  & 6&1 & 0&2  & 10&8 & 0&4  & 10&8 & 0&4  & 4&8 & 0&3 \\ 
 2&0 & 2&5  & 0&9 & 0&0  & 6&7 & 0&2  & 11&6 & 0&4  & 11&4 & 0&5  & 5&0 & 0&3 \\ 
 2&5 & 3&0  & 1&0 & 0&0  & 7&2 & 0&3  & 12&3 & 0&4  & 12&1 & 0&5  & 5&6 & 0&3 \\ 
 3&0 & 3&5  & 1&1 & 0&1  & 7&9 & 0&3  & 13&3 & 0&5  & 12&7 & 0&5  & 5&8 & 0&3 \\ 
 3&5 & 4&0  & 1&2 & 0&1  & 8&3 & 0&3  & 13&8 & 0&4  & 13&3 & 0&5  & 5&7 & 0&3 \\ 
 4&0 & 4&5  & 1&4 & 0&1  & 8&8 & 0&3  & 14&6 & 0&5  & 14&0 & 0&5  & 6&2 & 0&3 \\ 
 4&5 & 5&0  & 1&6 & 0&1  & 10&0 & 0&3  & 15&4 & 0&4  & 14&7 & 0&5  & 6&4 & 0&3 \\ 
 5&0 & 5&5  & 1&6 & 0&1  & 10&6 & 0&3  & 16&0 & 0&4  & 14&9 & 0&5  & 6&9 & 0&3 \\ 
 5&5 & 6&0  & 2&0 & 0&1  & 11&4 & 0&3  & 16&8 & 0&5  & 15&9 & 0&5  & 7&9 & 0&4 \\ 
 6&0 & 6&5  & 2&2 & 0&1  & 12&3 & 0&3  & 17&8 & 0&5  & 16&7 & 0&5  & 7&8 & 0&4 \\ 
 6&5 & 7&0  & 2&6 & 0&1  & 13&0 & 0&3  & 18&8 & 0&5  & 17&1 & 0&5  & 8&2 & 0&4 \\ 
 7&0 & 7&5  & 3&0 & 0&1  & 14&2 & 0&3  & 19&2 & 0&4  & 17&7 & 0&5  & 9&0 & 0&4 \\ 
 7&5 & 8&0  & 3&3 & 0&1  & 15&0 & 0&3  & 20&9 & 0&4  & 18&4 & 0&4  & 9&0 & 0&4 \\ 
 8&0 & 8&5  & 3&9 & 0&1  & 16&5 & 0&3  & 21&1 & 0&3  & 20&6 & 0&4  & 9&6 & 0&4 \\ 
 8&5 & 9&0  & 4&2 & 0&1  & 17&1 & 0&3  & 21&7 & 0&3  & 21&1 & 0&4  & 10&2 & 0&4 \\ 
 9&0 & 9&5  & 4&5 & 0&2  & 17&9 & 0&3  & 23&2 & 0&3  & 21&2 & 0&4  & 11&5 & 0&5 \\ 
 9&5 & 10&0  & 5&3 & 0&2  & 19&4 & 0&3  & 24&0 & 0&4  & 21&6 & 0&5  & 12&1 & 0&5 \\ 
 10&0 & 10&5  & 5&2 & 0&2  & 20&3 & 0&4  & 24&7 & 0&4  & 22&9 & 0&5  & 11&9 & 0&6 \\ 
 10&5 & 11&5  & 6&0 & 0&2  & 21&3 & 0&3  & 25&9 & 0&3  & 24&2 & 0&4  & 13&8 & 0&5 \\ 
 11&5 & 12&5  & 7&0 & 0&2  & 23&5 & 0&3  & 26&8 & 0&4  & 26&1 & 0&5  & 15&3 & 0&7 \\ 
 12&5 & 14&0  & 8&4 & 0&2  & 24&4 & 0&3  & 27&9 & 0&4  & 26&2 & 0&5  & 15&2 & 0&7 \\ 
 14&0 & 16&5  & 9&9 & 0&3  & 26&3 & 0&4  & 29&1 & 0&4  & 27&4 & 0&6  & 16&6 & 0&8 \\ 
 16&5 & 23&5  & 13&2 & 0&3  & 28&4 & 0&4  & 30&3 & 0&4  & 29&8 & 0&7  & 19&3 & 1&0 \\ 
 23&5 & 40&0  & 18&9 & 0&7  & 30&6 & 0&7  & 31&0 & 0&9  & 28&0 & 1&6  & 18&0 & 3&1 \\ 
\end{tabular}
\end{center}
\end{table}

\begin{table}[htbp]
\begin{center}
\caption{\label{tab:eff_13} Measured $\varepsilon_{\text{tot}}$ of \Bpm events at $13\protect\tev$,
in the bins of \Bpm \pt and $y$. The efficiencies and uncertainties are in percent.}
\begin{tabular}{r@{.}l@{$~-~$}r@{.}lr@{.}l@{$\pm$}r@{.}lr@{.}l@{$\pm$}r@{.}lr@{.}l@{$\pm$}r@{.}lr@{.}l@{$\pm$}r@{.}l r@{.}l@{$\pm$}r@{.}l}
\multicolumn{4}{c}{\pt [\gevc]} & \multicolumn{4}{c}{$2.0<y<2.5$} & \multicolumn{4}{c}{$2.5<y<3.0$} & \multicolumn{4}{c}{$3.0<y<3.5$} & \multicolumn{4}{c}{$3.5<y<4.0$} & \multicolumn{4}{c}{$4.0<y<4.5$}\\ \hline
 0&0 & 0&5  & 0&8 & 0&1   & 4&7 & 0&1   & 7&6 & 0&2   & 6&8 & 0&2   & 2&3 & 0&1  \\ 
 0&5 & 1&0  & 0&8 & 0&0   & 4&9 & 0&1   & 7&8 & 0&1   & 7&0 & 0&1   & 2&6 & 0&1  \\ 
 1&0 & 1&5  & 0&8 & 0&0   & 5&4 & 0&1   & 8&4 & 0&1   & 7&3 & 0&1   & 2&7 & 0&1  \\ 
 1&5 & 2&0  & 0&9 & 0&0   & 5&5 & 0&1   & 8&6 & 0&1   & 7&5 & 0&1   & 2&9 & 0&1  \\ 
 2&0 & 2&5  & 0&9 & 0&0   & 6&1 & 0&1   & 9&6 & 0&1   & 8&2 & 0&1   & 3&1 & 0&1  \\ 
 2&5 & 3&0  & 1&0 & 0&0   & 6&6 & 0&1   & 10&1 & 0&1   & 8&7 & 0&1   & 3&4 & 0&1  \\ 
 3&0 & 3&5  & 1&2 & 0&0   & 7&0 & 0&1   & 10&5 & 0&1   & 9&2 & 0&1   & 3&5 & 0&1  \\ 
 3&5 & 4&0  & 1&3 & 0&0   & 7&7 & 0&1   & 11&3 & 0&2   & 9&6 & 0&2   & 3&8 & 0&1  \\ 
 4&0 & 4&5  & 1&4 & 0&0   & 8&3 & 0&1   & 11&8 & 0&2   & 9&9 & 0&2   & 4&0 & 0&1  \\ 
 4&5 & 5&0  & 1&6 & 0&0   & 8&7 & 0&1   & 12&4 & 0&2   & 10&6 & 0&2   & 4&3 & 0&1  \\ 
 5&0 & 5&5  & 1&8 & 0&0   & 9&7 & 0&1   & 13&0 & 0&1   & 11&1 & 0&1   & 4&4 & 0&1  \\ 
 5&5 & 6&0  & 2&1 & 0&1   & 10&3 & 0&1   & 14&0 & 0&2   & 11&7 & 0&2   & 4&9 & 0&1  \\ 
 6&0 & 6&5  & 2&3 & 0&1   & 11&1 & 0&1   & 14&6 & 0&2   & 12&3 & 0&2   & 5&1 & 0&1  \\ 
 6&5 & 7&0  & 2&6 & 0&1   & 12&0 & 0&2   & 14&9 & 0&2   & 13&0 & 0&2   & 5&4 & 0&1  \\ 
 7&0 & 7&5  & 3&0 & 0&1   & 13&0 & 0&2   & 15&6 & 0&2   & 13&4 & 0&2   & 5&9 & 0&2  \\ 
 7&5 & 8&0  & 3&2 & 0&1   & 13&3 & 0&2   & 16&8 & 0&2   & 14&4 & 0&2   & 6&3 & 0&2  \\ 
 8&0 & 8&5  & 3&5 & 0&1   & 14&3 & 0&2   & 17&2 & 0&2   & 15&0 & 0&3   & 6&6 & 0&2  \\ 
 8&5 & 9&0  & 4&0 & 0&1   & 15&3 & 0&2   & 18&0 & 0&2   & 16&0 & 0&3   & 7&0 & 0&2  \\ 
 9&0 & 9&5  & 4&4 & 0&1   & 16&2 & 0&2   & 18&9 & 0&3   & 16&8 & 0&3   & 7&7 & 0&3  \\ 
 9&5 & 10&0  & 5&2 & 0&2   & 17&8 & 0&3   & 19&8 & 0&3   & 17&4 & 0&3   & 7&7 & 0&3  \\ 
 10&0 & 10&5  & 5&8 & 0&2   & 18&2 & 0&3   & 20&4 & 0&3   & 18&3 & 0&4   & 8&4 & 0&3  \\ 
 10&5 & 11&5  & 6&3 & 0&2   & 19&0 & 0&2   & 21&5 & 0&3   & 18&9 & 0&3   & 9&1 & 0&3  \\ 
 11&5 & 12&5  & 7&1 & 0&2   & 20&5 & 0&3   & 22&0 & 0&3   & 21&0 & 0&4   & 9&7 & 0&3  \\ 
 12&5 & 14&0  & 8&7 & 0&2   & 22&7 & 0&3   & 23&6 & 0&3   & 21&0 & 0&4   & 10&7 & 0&3  \\ 
 14&0 & 16&5  & 10&5 & 0&2   & 24&3 & 0&3   & 24&6 & 0&3   & 22&5 & 0&4   & 11&7 & 0&4  \\ 
 16&5 & 23&5  & 13&4 & 0&2   & 27&0 & 0&3   & 26&2 & 0&3   & 24&1 & 0&4   & 13&0 & 0&4  \\ 
 23&5 & 40&0  & 18&2 & 0&5   & 29&0 & 0&5   & 26&9 & 0&6   & 24&0 & 0&8   & 14&0 & 1&0  \\ 
\end{tabular}
\end{center}
\end{table}

\section{Tabulated results}
\label{sec:Supplementary-App}

The measured \Bpm double-differential cross-section
in bins of \pt and $y$ is given in
Tables~\ref{tab:2dxsec7} and~\ref{tab:2dxsec17} for 7\tev data,
and Tables~\ref{tab:2dxsec} and~\ref{tab:2dxsec1} for 13\tev data.
The measured \Bpm single-differential cross-section as a function of \pt or $y$ is given in Tables~\ref{tab:res_pt} and~\ref{tab:res_y}, respectively.
The limited size of the simulation samples gives relative systematic uncertainties in the high \pt region that are larger than those in the low \pt region. 

\begin{table}[htbp]
\begin{center}
\caption{\label{tab:2dxsec7} Measured \Bpm double-differential cross-section (in units of \nb) at $7\protect\tev$,
as a function of \pt and $y$, in the rapidity regions of $2.0<y<2.5$, $2.5<y<3.0$, and $3.0<y<3.5$.}
\begin{tabular}{r@{.}l@{$~-~$}r@{.}lr@{.}l@{$\pm$}r@{.}l@{$\pm$}r@{.}lr@{.}l@{$\pm$}r@{.}l@{$\pm$}r@{.}lr@{.}l@{$\pm$}r@{.}l@{$\pm$}r@{.}l}
\multicolumn{4}{c}{\pt [\gevc]} & \multicolumn{6}{c}{$2.0<y<2.5$} & \multicolumn{6}{c}{$2.5<y<3.0$} & \multicolumn{6}{c}{$3.0<y<3.5$} \\ \hline
  0&0&  0&5  &  664&5 & 103&3 & 99&9 &  519&8 & 35&0 & 50&3 &  462&6 & 30&4 & 37&5 \\ 
  0&5&  1&0  &  1652&9 & 160&9 & 180&6 &  1544&9 & 59&8 & 142&0 &  1248&8 & 51&6 & 94&4 \\ 
  1&0&  1&5  &  2204&8 & 171&7 & 345&8 &  2381&1 & 71&0 & 212&8 &  2046&2 & 62&4 & 147&1 \\ 
  1&5&  2&0  &  2879&5 & 191&3 & 311&6 &  3175&6 & 76&2 & 260&1 &  2665&7 & 65&1 & 192&5 \\ 
  2&0&  2&5  &  3378&9 & 192&0 & 592&5 &  3193&0 & 80&1 & 257&7 &  3017&2 & 64&1 & 214&6 \\ 
  2&5&  3&0  &  3261&7 & 179&8 & 321&9 &  3605&8 & 69&9 & 256&6 &  3056&6 & 60&7 & 220&0 \\ 
  3&0&  3&5  &  3627&8 & 182&8 & 290&3 &  3263&8 & 66&6 & 263&4 &  2977&6 & 54&6 & 194&2 \\ 
  3&5&  4&0  &  3491&7 & 175&6 & 419&5 &  3304&0 & 57&9 & 246&3 &  2767&2 & 49&8 & 167&7 \\ 
  4&0&  4&5  &  2976&6 & 146&4 & 240&0 &  3224&0 & 50&9 & 233&4 &  2545&6 & 44&7 & 173&3 \\ 
  4&5&  5&0  &  3216&9 & 141&1 & 255&9 &  2773&2 & 44&1 & 185&0 &  2258&6 & 39&7 & 134&8 \\ 
  5&0&  5&5  &  2795&0 & 127&6 & 308&6 &  2408&0 & 41&3 & 158&0 &  2063&8 & 36&1 & 128&5 \\ 
  5&5&  6&0  &  2183&8 & 101&3 & 160&2 &  2145&3 & 34&4 & 136&9 &  1740&5 & 31&3 & 108&4 \\ 
  6&0&  6&5  &  2194&2 & 94&8 & 200&1 &  1834&6 & 28&8 & 124&7 &  1437&9 & 27&3 & 85&9 \\ 
  6&5&  7&0  &  1694&1 & 75&0 & 166&0 &  1645&1 & 24&4 & 109&0 &  1219&0 & 23&7 & 69&7 \\ 
  7&0&  7&5  &  1517&7 & 65&6 & 102&6 &  1329&8 & 21&9 & 84&9 &  1039&8 & 21&4 & 59&3 \\ 
  7&5&  8&0  &  1200&0 & 54&2 & 134&2 &  1101&1 & 18&7 & 73&8 &  842&3 & 18&1 & 48&4 \\ 
  8&0&  8&5  &  1036&9 & 46&2 & 91&2 &  950&4 & 16&7 & 55&4 &  744&6 & 16&7 & 41&3 \\ 
  8&5&  9&0  &  943&1 & 42&4 & 124&1 &  851&6 & 14&1 & 50&5 &  614&3 & 14&9 & 33&3 \\ 
  9&0&  9&5  &  782&0 & 36&7 & 57&8 &  687&3 & 13&9 & 39&6 &  554&8 & 13&5 & 30&0 \\ 
  9&5&  10&0  &  617&0 & 30&0 & 49&2 &  559&9 & 11&9 & 32&8 &  448&5 & 11&9 & 27&1 \\ 
  10&0&  10&5  &  594&6 & 29&4 & 79&9 &  487&7 & 10&2 & 26&9 &  371&0 & 10&7 & 21&6 \\ 
  10&5&  11&5  &  502&1 & 17&6 & 55&1 &  409&9 & 5&9 & 25&7 &  287&9 & 6&4 & 15&3 \\ 
  11&5&  12&5  &  349&3 & 13&5 & 38&7 &  278&0 & 5&0 & 14&7 &  212&3 & 5&4 & 11&6 \\ 
  12&5&  14&0  &  241&5 & 8&4 & 20&0 &  204&5 & 3&1 & 11&5 &  142&1 & 3&5 & 9&9 \\ 
  14&0&  16&5  &  146&9 & 4&6 & 9&2 &  115&4 & 1&7 & 6&3 &  79&2 & 2&0 & 6&5 \\ 
  16&5&  23&5  &  49&9 & 1&4 & 2&9 &  40&0 & 0&5 & 2&9 &  24&8 & 0&6 & 1&9 \\ 
  23&5&  40&0  &  6&1 & 0&3 & 0&4 &  5&1 & 0&1 & 0&4 &  2&7 & 0&1 & 0&3 \\ 
\end{tabular}
\end{center}
\end{table}

\begin{table}[htbp]
\begin{center}
\caption{\label{tab:2dxsec17} Measured \Bpm double-differential cross-section (in units of \nb) at $7\protect\tev$,
as a function of \pt and $y$, in the rapidity regions of $3.5<y<4.0$ and $4.0<y<4.5$.}
\begin{tabular}{r@{.}l@{$~-~$}r@{.}lr@{.}l@{$\pm$}r@{.}l@{$\pm$}r@{.}lr@{.}l@{$\pm$}r@{.}l@{$\pm$}r@{.}l}
\multicolumn{4}{c}{\pt (\gevc)} & \multicolumn{6}{c}{$3.5<y<4.0$} & \multicolumn{6}{c}{$4.0<y<4.5$} \\ \hline
  0&0 & 0&5  & 396&4 & 29&2 & 34&4 & 283&7 & 37&9 & 35&3 \\ 
  0&5 & 1&0  & 1069&6 & 43&9 & 83&6 & 820&5 & 57&0 & 83&9 \\ 
  1&0 & 1&5  & 1581&2 & 50&0 & 121&1 & 1115&7 & 63&6 & 102&2 \\ 
  1&5 & 2&0  & 2132&1 & 53&8 & 161&0 & 1447&6 & 65&7 & 146&7 \\ 
  2&0 & 2&5  & 2256&6 & 52&1 & 165&1 & 1570&9 & 63&2 & 145&3 \\ 
  2&5 & 3&0  & 2241&9 & 48&6 & 158&9 & 1627&4 & 58&5 & 133&5 \\ 
  3&0 & 3&5  & 2265&2 & 45&6 & 157&5 & 1566&6 & 55&1 & 134&3 \\ 
  3&5 & 4&0  & 2094&9 & 42&2 & 147&0 & 1465&1 & 52&1 & 113&5 \\ 
  4&0 & 4&5  & 2002&0 & 39&2 & 133&4 & 1259&9 & 45&5 & 114&5 \\ 
  4&5 & 5&0  & 1642&0 & 34&0 & 101&1 & 1144&0 & 42&8 & 101&4 \\ 
  5&0 & 5&5  & 1569&5 & 32&0 & 108&9 & 961&9 & 37&0 & 76&5 \\ 
  5&5 & 6&0  & 1223&2 & 27&2 & 75&1 & 734&2 & 29&8 & 73&7 \\ 
  6&0 & 6&5  & 1038&0 & 23&7 & 62&5 & 652&5 & 27&6 & 48&4 \\ 
  6&5 & 7&0  & 861&0 & 20&8 & 53&3 & 548&6 & 24&1 & 44&0 \\ 
  7&0 & 7&5  & 704&6 & 18&2 & 40&9 & 390&1 & 19&2 & 28&5 \\ 
  7&5 & 8&0  & 628&7 & 16&6 & 37&2 & 326&4 & 17&4 & 27&7 \\ 
  8&0 & 8&5  & 465&3 & 13&4 & 28&3 & 280&7 & 15&3 & 21&2 \\ 
  8&5 & 9&0  & 403&4 & 12&2 & 22&3 & 241&0 & 13&5 & 17&7 \\ 
  9&0 & 9&5  & 330&0 & 11&0 & 20&8 & 190&6 & 11&4 & 17&6 \\ 
  9&5 & 10&0  & 286&9 & 10&2 & 17&7 & 163&9 & 10&2 & 14&2 \\ 
  10&0&  10&5  & 230&0 & 8&8 & 13&8 & 136&7 & 9&5 & 15&3 \\ 
  10&5&  11&5  & 186&7 & 5&4 & 10&5 & 87&0 & 4&9 & 8&2 \\ 
  11&5&  12&5  & 127&7 & 4&3 & 7&3 & 65&0 & 4&0 & 5&5 \\ 
  12&5&  14&0  & 84&4 & 2&8 & 4&8 & 40&4 & 2&6 & 3&5 \\ 
  14&0&  16&5  & 44&7 & 1&6 & 2&7 & 21&8 & 1&4 & 1&8 \\ 
  16&5&  23&5  & 12&4 & 0&5 & 0&8 & 4&4 & 0&3 & 0&4 \\ 
  23&5&  40&0  & 1&0 & 0&1 & 0&1 & 0&3 & 0&1 & 0&1 \\ 
\end{tabular}
\end{center}
\end{table}

\begin{table}[htbp]
\begin{center}
\caption{\label{tab:2dxsec} Measured \Bpm double-differential cross-section (in units of \nb) at $13\protect\tev$,
as a function of \pt and $y$, in the rapidity regions of $2.0<y<2.5$, $2.5<y<3.0$, and $3.0<y<3.5$.}
\begin{tabular}{r@{.}l@{$~-~$}r@{.}lr@{.}l@{$\pm$}r@{.}l@{$\pm$}r@{.}lr@{.}l@{$\pm$}r@{.}l@{$\pm$}r@{.}lr@{.}l@{$\pm$}r@{.}l@{$\pm$}r@{.}l}
\multicolumn{4}{c}{\pt [\gevc]} & \multicolumn{6}{c}{$2.0<y<2.5$} & \multicolumn{6}{c}{$2.5<y<3.0$} & \multicolumn{6}{c}{$3.0<y<3.5$} \\ \hline

  0&0 & 0&5  &  794&4 & 228&1 & 130&0 &  905&7 & 152&4 & 85&6 &  1013&9 & 183&2 & 81&1 \\ 
  0&5 & 1&0  &  2384&0 & 362&6 & 322&3 &  3371&6 & 178&9 & 267&2 &  2221&2 & 194&6 & 152&5 \\ 
  1&0 & 1&5  &  4503&0 & 493&3 & 419&1 &  4211&5 & 232&8 & 321&1 &  3557&8 & 223&2 & 236&1 \\ 
  1&5 & 2&0  &  6378&3 & 557&8 & 939&8 &  4846&7 & 297&1 & 356&4 &  5255&5 & 236&4 & 370&7 \\ 
  2&0 & 2&5  &  5543&9 & 518&5 & 501&1 &  6186&6 & 236&4 & 465&9 &  5136&2 & 214&9 & 347&9 \\ 
  2&5 & 3&0  &  7517&2 & 555&3 & 695&2 &  6739&7 & 217&1 & 477&0 &  5519&8 & 203&8 & 369&2 \\ 
  3&0 & 3&5  &  6848&8 & 534&6 & 738&4 &  6152&6 & 228&5 & 412&6 &  5903&7 & 189&0 & 419&3 \\ 
  3&5 & 4&0  &  6382&6 & 475&5 & 509&5 &  6321&5 & 187&5 & 452&4 &  5560&9 & 167&3 & 368&1 \\ 
  4&0 & 4&5  &  5900&1 & 449&8 & 543&6 &  5525&8 & 162&8 & 371&3 &  4824&6 & 144&4 & 315&7 \\ 
  4&5 & 5&0  &  6185&4 & 412&5 & 503&8 &  5259&5 & 150&0 & 368&0 &  4657&9 & 131&8 & 318&8 \\ 
  5&0 & 5&5  &  5353&0 & 349&5 & 402&2 &  4735&5 & 119&4 & 313&3 &  3900&2 & 113&6 & 256&2 \\ 
  5&5 & 6&0  &  4168&6 & 279&7 & 357&4 &  4300&4 & 105&7 & 290&2 &  3535&4 & 101&0 & 245&1 \\ 
  6&0 & 6&5  &  3532&0 & 237&5 & 326&4 &  3865&7 & 92&1 & 256&5 &  3158&4 & 88&8 & 211&9 \\ 
  6&5 & 7&0  &  3111&7 & 203&9 & 259&2 &  3107&7 & 77&0 & 227&6 &  2544&5 & 77&0 & 167&0 \\ 
  7&0 & 7&5  &  2891&4 & 174&3 & 282&1 &  2662&1 & 63&4 & 177&2 &  2086&4 & 66&5 & 143&7 \\ 
  7&5 & 8&0  &  2270&4 & 149&5 & 182&8 &  2450&3 & 54&9 & 164&1 &  1788&4 & 58&2 & 122&9 \\ 
  8&0 & 8&5  &  2467&8 & 143&9 & 213&3 &  2034&0 & 52&1 & 149&6 &  1638&5 & 53&4 & 109&7 \\ 
  8&5 & 9&0  &  1947&3 & 118&9 & 195&0 &  1710&1 & 44&0 & 118&7 &  1320&5 & 46&2 & 87&5 \\ 
  9&0 & 9&5  &  1424&4 & 98&0 & 115&2 &  1472&0 & 41&9 & 99&1 &  1227&2 & 42&8 & 83&0 \\ 
  9&5 & 10&0  &  1283&7 & 84&4 & 96&4 &  1243&8 & 36&7 & 86&2 &  1039&0 & 38&0 & 74&6 \\ 
  10&0 & 10&5  &  1313&6 & 78&7 & 100&9 &  1189&3 & 31&3 & 79&5 &  883&8 & 34&3 & 59&3 \\ 
  10&5 & 11&5  &  870&5 & 43&3 & 67&7 &  936&9 & 18&1 & 68&1 &  666&8 & 20&5 & 44&4 \\ 
  11&5 & 12&5  &  802&2 & 38&1 & 66&1 &  632&3 & 17&1 & 42&4 &  541&7 & 17&9 & 36&4 \\ 
  12&5 & 14&0  &  518&5 & 22&4 & 36&2 &  487&1 & 10&1 & 32&1 &  365&3 & 11&6 & 24&7 \\ 
  14&0 & 16&5  &  333&0 & 12&5 & 23&2 &  289&2 & 5&8 & 19&6 &  216&4 & 6&7 & 14&3 \\ 
  16&5 & 23&5  &  123&1 & 4&0 & 9&3 &  101&0 & 1&8 & 6&6 &  72&4 & 2&2 & 5&1 \\ 
  23&5 & 40&0  &  19&9 & 0&9 & 1&4 &  14&3 & 0&4 & 1&1 &  9&8 & 0&5 & 0&7 \\ 
\end{tabular}
\end{center}
\end{table}

\begin{table}[htbp]
\begin{center}
\caption{\label{tab:2dxsec1} Measured \Bpm double-differential cross-section (in units of \nb) at $13\protect\tev$,
as a function of \pt and $y$, in the rapidity regions of $3.5<y<4.0$ and $4.0<y<4.5$.}
\begin{tabular}{r@{.}l@{$~-~$}r@{.}lr@{.}l@{$\pm$}r@{.}l@{$\pm$}r@{.}lr@{.}l@{$\pm$}r@{.}l@{$\pm$}r@{.}l}
\multicolumn{4}{c}{\pt (\gevc)} & \multicolumn{6}{c}{$3.5<y<4.0$} & \multicolumn{6}{c}{$4.0<y<4.5$} \\ \hline

  0&0 & 0&5  & 814&7 & 111&2 & 83&4 & 369&7 & 156&1 & 48&4 \\ 
  0&5 & 1&0  & 2037&6 & 161&1 & 165&9 & 1235&7 & 197&9 & 107&4 \\ 
  1&0 & 1&5  & 3120&6 & 186&1 & 288&9 & 2649&3 & 249&5 & 236&6 \\ 
  1&5 & 2&0  & 3980&7 & 202&0 & 266&5 & 2605&7 & 236&3 & 238&8 \\ 
  2&0 & 2&5  & 4187&1 & 185&7 & 288&7 & 3079&7 & 240&3 & 232&1 \\ 
  2&5 & 3&0  & 4869&2 & 179&4 & 332&8 & 3237&0 & 224&2 & 252&6 \\ 
  3&0 & 3&5  & 4898&3 & 168&1 & 327&0 & 3246&0 & 210&9 & 257&8 \\ 
  3&5 & 4&0  & 4336&9 & 149&4 & 343&3 & 2807&4 & 185&1 & 268&8 \\ 
  4&0 & 4&5  & 4118&0 & 140&2 & 283&3 & 2804&3 & 171&0 & 225&4 \\ 
  4&5 & 5&0  & 3690&8 & 121&8 & 249&0 & 2611&4 & 153&9 & 187&4 \\ 
  5&0 & 5&5  & 3248&0 & 108&5 & 213&2 & 2013&2 & 133&7 & 151&5 \\ 
  5&5 & 6&0  & 2910&9 & 96&5 & 193&3 & 1886&3 & 123&0 & 207&7 \\ 
  6&0 & 6&5  & 2381&8 & 83&6 & 157&1 & 1610&8 & 105&5 & 129&0 \\ 
  6&5 & 7&0  & 1907&1 & 70&2 & 126&9 & 1415&9 & 93&2 & 112&7 \\ 
  7&0 & 7&5  & 1661&2 & 63&4 & 119&0 & 1095&4 & 76&5 & 83&1 \\ 
  7&5 & 8&0  & 1364&5 & 54&1 & 96&8 & 910&1 & 66&1 & 74&4 \\ 
  8&0 & 8&5  & 1238&8 & 49&7 & 83&7 & 826&1 & 60&3 & 63&9 \\ 
  8&5 & 9&0  & 961&2 & 41&8 & 64&3 & 581&5 & 50&7 & 50&9 \\ 
  9&0 & 9&5  & 848&9 & 38&4 & 57&0 & 528&5 & 44&7 & 43&7 \\ 
  9&5 & 10&0  & 732&8 & 34&6 & 50&4 & 518&6 & 43&0 & 43&7 \\ 
  10&0 & 10&5  & 635&7 & 30&9 & 60&2 & 350&2 & 33&9 & 32&1 \\ 
  10&5 & 11&5  & 474&4 & 18&4 & 33&1 & 308&7 & 21&6 & 25&2 \\ 
  11&5 & 12&5  & 343&0 & 14&7 & 24&8 & 235&5 & 18&0 & 19&1 \\ 
  12&5 & 14&0  & 242&6 & 10&2 & 16&3 & 147&9 & 11&1 & 13&1 \\ 
  14&0 & 16&5  & 131&0 & 5&6 & 8&8 & 71&8 & 5&7 & 5&7 \\ 
  16&5 & 23&5  & 45&9 & 1&9 & 3&2 & 22&8 & 1&9 & 2&0 \\ 
  23&5 & 40&0  & 5&2 & 0&4 & 0&4 & 2&3 & 0&4 & 0&3 \\ 
\end{tabular}
\end{center}
\end{table}

\begin{table}[htbp]
  \begin{center}
    \caption{\label{tab:res_pt} Measured \Bpm differential cross-sections (in units of $\nb$) at
$7\protect\tev$ and $13\protect\tev$ as functions of \pt in the range $2.0<y<4.5$.
The cross-section ratio between $13\protect\tev$ and $7\protect\tev$ is also presented.}
     \begin{tabular}{r@{.}l@{$~-~$}r@{.}lr@{.}l@{$\pm$}r@{.}l@{$\pm$}r@{.}lr@{.}l@{$\pm$}r@{.}l@{$\pm$}r@{.}lr@{.}l@{$\pm$}r@{.}l@{$\pm$}r@{.}l}
    \multicolumn{4}{c}{\pt [\gevc]}& \multicolumn{6}{c}{$7\protect\tev$} & \multicolumn{6}{c}{$13\protect\tev$}  & \multicolumn{6}{c}{$R(13/7)$}\\ \hline
0&0&0&5 & 1163&5 & 62&1 & 101&8 & 1949&2 & 187&7 & 182&4 & 1&68 & 0&18 & 0&17 \\ 
 0&5&1&0 & 3168&4 & 99&0 & 237&0 & 5625&1 & 277&6 & 444&3 & 1&78 & 0&10 & 0&14 \\ 
 1&0&1&5 & 4664&6 & 108&1 & 383&1 & 9021&1 & 341&3 & 676&1 & 1&93 & 0&09 & 0&16 \\ 
 1&5&2&0 & 6150&2 & 118&6 & 459&3 & 11533&4 & 367&0 & 972&4 & 1&88 & 0&07 & 0&16 \\ 
 2&0&2&5 & 6708&3 & 117&1 & 566&0 & 12066&8 & 349&4 & 856&8 & 1&80 & 0&06 & 0&15 \\ 
 2&5&3&0 & 6896&7 & 110&2 & 454&0 & 13941&5 & 354&5 & 978&2 & 2&02 & 0&06 & 0&13 \\ 
 3&0&3&5 & 6850&5 & 108&2 & 442&2 & 13524&7 & 335&7 & 981&9 & 1&97 & 0&06 & 0&13 \\ 
 3&5&4&0 & 6561&5 & 103&2 & 462&6 & 12704&7 & 298&4 & 881&2 & 1&94 & 0&05 & 0&14 \\ 
 4&0&4&5 & 6004&1 & 88&3 & 388&6 & 11586&5 & 277&0 & 794&5 & 1&93 & 0&05 & 0&12 \\ 
 4&5&5&0 & 5517&3 & 82&8 & 338&2 & 11202&5 & 252&3 & 761&9 & 2&03 & 0&05 & 0&12 \\ 
 5&0&5&5 & 4899&1 & 74&7 & 335&9 & 9625&0 & 215&4 & 628&0 & 1&96 & 0&05 & 0&13 \\ 
 5&5&6&0 & 4013&5 & 60&5 & 238&5 & 8400&8 & 179&9 & 594&3 & 2&09 & 0&05 & 0&13 \\ 
 6&0&6&5 & 3578&6 & 55&7 & 222&7 & 7274&4 & 154&2 & 494&3 & 2&03 & 0&05 & 0&13 \\ 
 6&5&7&0 & 2983&9 & 45&5 & 189&6 & 6043&4 & 132&2 & 415&3 & 2&03 & 0&05 & 0&13 \\ 
 7&0&7&5 & 2491&0 & 39&5 & 137&1 & 5198&3 & 113&2 & 369&3 & 2&09 & 0&06 & 0&13 \\ 
 7&5&8&0 & 2049&2 & 33&3 & 141&3 & 4391&8 & 98&3 & 297&9 & 2&14 & 0&06 & 0&15 \\ 
 8&0&8&5 & 1738&9 & 28&6 & 104&5 & 4102&6 & 92&0 & 286&4 & 2&36 & 0&07 & 0&15 \\ 
 8&5&9&0 & 1526&7 & 26&2 & 104&5 & 3260&3 & 77&2 & 233&3 & 2&14 & 0&06 & 0&16 \\ 
 9&0&9&5 & 1272&3 & 22&8 & 73&1 & 2750&5 & 66&0 & 183&0 & 2&16 & 0&06 & 0&13 \\ 
 9&5&10&0 & 1038&1 & 19&2 & 62&6 & 2408&9 & 58&2 & 160&6 & 2&32 & 0&07 & 0&14 \\ 
 10&0&10&5 & 909&9 & 18&2 & 65&8 & 2186&3 & 53&0 & 148&1 & 2&40 & 0&08 & 0&18 \\ 
 10&5&11&5 & 736&8 & 10&9 & 51&2 & 1628&7 & 30&6 & 113&0 & 2&21 & 0&05 & 0&16 \\ 
 11&5&12&5 & 516&2 & 8&5 & 32&7 & 1277&4 & 26&0 & 87&0 & 2&47 & 0&06 & 0&16 \\ 
 12&5&14&0 & 356&5 & 5&4 & 22&3 & 880&7 & 16&1 & 57&2 & 2&47 & 0&06 & 0&15 \\ 
 14&0&16&5 & 204&0 & 3&0 & 11&8 & 520&7 & 9&0 & 33&9 & 2&55 & 0&06 & 0&15 \\ 
 16&5&23&5 & 65&8 & 0&9 & 3&9 & 182&6 & 2&9 & 12&4 & 2&78 & 0&06 & 0&17 \\ 
 23&5&40&0 & 7&6 & 0&2 & 0&6 & 25&7 & 0&6 & 1&8 & 3&38 & 0&12 & 0&25 \\ 
    \end{tabular}
  \end{center}
\end{table}

\begin{table}[htbp]
  \begin{center}
    \caption{\label{tab:res_y} Measured \Bpm differential cross-sections (in units of $\mub$) at
$7\protect\tev$ and $13\protect\tev$ as functions of $y$ in the \pt range $0<\pt<40\gevc$.
The cross-section ratio between $13\protect\tev$ and $7\protect\tev$ is also presented.}
    \begin{tabular}{r@{.}l@{$~-~$}r@{.}lr@{.}l@{$\pm$}r@{.}l@{$\pm$}r@{.}lr@{.}l@{$\pm$}r@{.}l@{$\pm$}r@{.}lr@{.}l@{$\pm$}r@{.}l@{$\pm$}r@{.}l}
   \multicolumn{3}{c}{ $y $} & \multicolumn{6}{c}{$7\protect\tev$} & \multicolumn{6}{c}{$13\protect\tev$}  & \multicolumn{6}{c}{$R(13/7)$}\\ \hline
 2&0&2&5 & 23&5 & 0&3 & 2&0 & 45&6 & 0&8 & 3&6 & 1&94 & 0&04 & 0&19 \\ 
 2&5&3&0 & 22&1 & 0&1 & 1&4 & 43&1 & 0&4 & 2&9 & 1&95 & 0&02 & 0&12 \\ 
 3&0&3&5 & 18&2 & 0&1 & 1&0 & 36&4 & 0&3 & 2&4 & 2&00 & 0&02 & 0&10 \\ 
 3&5&4&0 & 13&4 & 0&1 & 0&8 & 28&9 & 0&3 & 1&9 & 2&16 & 0&02 & 0&12 \\ 
 4&0&4&5 & 8&8 & 0&1 & 0&6 & 19&3 & 0&4 & 1&4 & 2&21 & 0&05 & 0&16 \\ 
    \end{tabular}
  \end{center}
\end{table}

\clearpage

%% file: LHCb_Authorship_flat_10-Aug-2017.tex
\centerline{\large\bf LHCb collaboration}
\begin{flushleft}
\small
R.~Aaij$^{40}$,
B.~Adeva$^{39}$,
M.~Adinolfi$^{48}$,
Z.~Ajaltouni$^{5}$,
S.~Akar$^{59}$,
J.~Albrecht$^{10}$,
F.~Alessio$^{40}$,
M.~Alexander$^{53}$,
A.~Alfonso~Albero$^{38}$,
S.~Ali$^{43}$
% ...

{\footnotesize \it
% ...
$ ^{5}$Clermont Universit{\'e}, Universit{\'e} Blaise Pascal, CNRS/IN2P3, LPC, Clermont-Ferrand, France\\
% ...
$ ^{10}$Fakult{\"a}t Physik, Technische Universit{\"a}t Dortmund, Dortmund, Germany\\
% ...
$ ^{38}$ICCUB, Universitat de Barcelona, Barcelona, Spain\\
$ ^{39}$Universidad de Santiago de Compostela, Santiago de Compostela, Spain\\
$ ^{40}$European Organization for Nuclear Research (CERN), Geneva, Switzerland\\
% ...
$ ^{43}$Nikhef National Institute for Subatomic Physics, Amsterdam, The Netherlands\\
% ...
$ ^{48}$H.H. Wills Physics Laboratory, University of Bristol, Bristol, United Kingdom\\
% ...
$ ^{53}$School of Physics and Astronomy, University of Glasgow, Glasgow, United Kingdom\\
% ...
$ ^{59}$University of Cincinnati, Cincinnati, OH, United States\\
% ...
}
\end{flushleft}